\documentclass[fleqn,usenatbib]{mnras}

\usepackage[T1]{fontenc}

\DeclareRobustCommand{\VAN}[3]{#2}
\let\VANthebibliography\thebibliography
\def\thebibliography{\DeclareRobustCommand{\VAN}[3]{##3}\VANthebibliography}

\usepackage{graphicx}	

\usepackage{multicol}
\usepackage[export]{adjustbox}
\usepackage{multirow}
\usepackage{hyperref}
\usepackage{float}
\usepackage[strict]{changepage}
\usepackage{rotating}
\usepackage{eqnarray,amsmath}
\usepackage{footnote}
\usepackage[svgnames]{xcolor}
\usepackage{tabularx}
\usepackage{arydshln}
\usepackage{tikz}
\usepackage{hyperref}
\usepackage{geometry}
\usepackage{enumitem}
\usepackage{caption}
\usepackage{newtxtext,newtxmath}
\usepackage{textcomp}
\usepackage{subfigure}

\title[Cosmological simulations of momentum coupled DM-DE]{Cosmological simulations of a momentum coupling between dark matter and quintessence}

\author[D. Palma et al.]{
Daniela Palma,$^{1,2}$\thanks{E-mail: daniela.palma.21@alumnos.uda.cl}
Graeme N. Candlish$^{2}$\thanks{E-mail: graeme.candlish@uv.cl}
\\
$^{1}$Instituto de Astronom\'ia y Ciencias Planetarias de Atacama, Universidad de Atacama, Copayapu 485, Copiap\'o, Chile\\
$^{2}$Instituto de F\'isica y Astronom\'ia, Universidad de Valpara\'iso, Gran Breta\~na 1111, Valpara\'iso, Chile\\
}

\date{Accepted XXX. Received YYY; in original form ZZZ}

\pubyear{2021}

\begin{document}
\label{firstpage}
\pagerange{\pageref{firstpage}--\pageref{lastpage}}
\maketitle

\begin{abstract}
Dark energy is frequently modelled as an additional dynamical scalar field component in the Universe, referred to as ``quintessence,'' which drives the late-time acceleration. Furthermore, the quintessence field may be coupled to dark matter and/or baryons, leading to a fifth force. In this paper we explore the consequences for non-linear cosmological structure formation arising from a momentum coupling between the quintessence field and dark matter only. The coupling leads to a modified Euler equation, which we implement in an N-body cosmological simulation. We then analyse the effects of the coupling on the non-linear power spectrum and the properties of the dark matter halos. We find that, for certain quintessence potentials, a positive coupling can lead to significantly reduced structure on small scales and somewhat enhanced structure on large scales, as well as reduced halo density profiles and increased velocity dispersions.
\end{abstract}




\section{Introduction}
Cosmological observations \citep{planck2018} provide strong evidence that we live in a spatially flat universe presently undergoing an accelerated expansion, with the standard assumption being that this is driven by a cosmological constant, $\Lambda$. The other components of the standard $\Lambda$CDM model are dark matter (DM), whose only cosmologically-relevant interaction is through gravity, and the directly-observable baryonic matter of the visible Universe. This model has proven to be remarkably consistent with observations of the CMB, BAO and Hubble parameter, at least at high redshift \citep{Planck2013, Planck2015, planck2018}.

There are, however, several problems with the standard model (for a review see \citealp{2021challengesLCDM}). Firstly, while the cosmological constant may be interpreted as a homogeneous energy density that fills the Universe, referred to as dark energy (DE), the physical interpretation of this energy density from particle physics as arising from zero-point quantum fluctuations in the vacuum is famously problematic. 
Thus there are many proposals that the late-time accelerated expansion is caused by some other effect, either a modification of General Relativity or an additional dynamical dark energy component. Among the latter possibilities (see the review \citealp{Yoo_2012}), the most studied is a scalar field, first postulated in \cite{Wetterich_1995} and subsequently referred to as quintessence by \cite{Caldwell_1998}.

Another problem associated with DE in the form of a cosmological constant is the ``coincidence problem'' \citep{libro} which seems to imply that our current epoch is special, given that, in the standard model, the beginning of the DE dominated stage of the cosmological evolution occurred very recently, at a redshift of approximately $z \approx 0.3$. Dynamical DE theories, in particular those that include a coupling with DM, may help explain this apparent coincidence (\citealp{Amendola_1999}, although see \citealp{Lindner_2006} for an alternative viewpoint).

The tension between local and early-Universe measurements of the Hubble parameter is also a potentially serious issue with the standard model \citep{Verde}. While it may be that systematic errors in the measurements alleviate some of the tension \citep{Bernal_2016}, the size of the discrepancies suggests that it cannot be explained by assuming these errors alone.

In addition, the cosmological parameter $\sigma_8$, closely connected to the matter density parameter $\Omega_M$, provides us with an excellent tool to constrain matter formation and structure growth. The CMB and large-scale structure (LSS) observations have shown some discrepancies in this value, suggesting a tension between $\Omega_M$ and $\sigma_8$, as discussed in \cite{Planck_2013_sz}, indicating a lower structure growth rate than expected according to $\Lambda$CDM.

Furthermore, at small scales, several studies, such as that of \cite{Oh:2015xoa}, have shown through rotation curves that the density profiles of satellite galaxy halos exhibit an inner core of nearly constant density. This contrasts with predictions from numerical simulations using the $\Lambda$CDM model that halos have a universal NFW density profile, which shows a "cuspy" behaviour in the innermost regions, referred to as the \textit{cusp-core} problem. One possible solution comes from the contribution of feedback from the baryonic content of the halo, causing a redistribution of the dark matter and resulting in a core \citep{Valenzuela_2007}. It is not currently clear if such a mechanism is sufficient to resolve the problem, especially in very dark matter-dominated galaxies.

Another possible challenge to the model is the \textit{missing satellites} problem, whereby cosmological simulations suggest the existence of far more satellite dark matter halos than have been detected observationally (via their baryonic content) around the Milky Way (or other local-group galaxies). Various solutions have been proposed, such as reionisation suppressing star formation in low mass halos \citep{Bullock_2000} or tidal interactions stripping the baryonic material \citep{Brooks_2013}. Again, a clearer understanding of baryonic processes inside the halos may provide a solution within the context of $\Lambda$CDM.

Finally, the study of the motion of satellite galaxies around the Milky Way has shown discrepancies concerning $\Lambda$CDM predictions. The expectation within the hierarchical structure formation scenario is that there would be an approximately isotropic distribution of these galaxies around their hosts. However, as studied in the Milky Way \citep{Pawlowski_2013}, in Andromeda \citep{Koch_2006} and the elliptical galaxy Centaurus A \citep{M_ller_2018}, it appears that their satellite galaxies are positioned in a planar distribution around the host galaxy, around which they have a coherent rotational motion. This structure is referred to as the \textit{plane of satellites}. There are some indications that this problem may not be as severe for $\Lambda$CDM as initially thought, due to the non-isotropic build-up of structure falling in through filaments \citep{Zentner_2005}. However, given that observations suggest these satellite planes are ubiquitous \citep{Phillips_2015} it is still not at all clear if this can be generally accommodated within the standard model.

To address the large-scale problems associated to $\Lambda$ discussed above, and in particular the coincidence problem, there have been numerous proposals that the dynamical dark energy component may be coupled to the dark matter component. There is some motivation for this from particle physics, given that quantum corrections typically introduce couplings between particle species, and it is generally the \textit{absence} of a coupling that must be explained, usually via some postulated symmetry. By restricting the coupling to be between the two dark sector components we can also avoid very stringent fifth-force constraints \citep{fifth_force_study}.

The presence of such a coupling can strongly influence both the background evolution and that of the perturbations. In \cite{Amendola2000}, a model with linear coupling was studied, considering a quintessence scalar field with exponential potential. It was found that the effect of the coupling on the power spectrum reduced and increased (very slightly) the value of $\sigma_8$ for large and small couplings, respectively. 
In \cite{Valiviita_2008}, the authors studied a coupled model where the DM and DE components are treated as fluids. This type of coupling generated instabilities within the model, which, according to the same authors, could be avoided if the DE were considered as quintessence. \cite{Salvatelli_2014} proposed that if there is an interaction at the level of the densities, the observational data favour that it is activated in the later stages of the evolution of the Universe, $z \sim 0.9$. In the majority of studies, the coupling is usually introduced at the level of the continuity equations for DM and DE. This type of coupling  modifies the background evolution, so the coupling parameter must be very small \citep{Wang2016}, in order not to deviate excessively from the $\Lambda$CDM background predictions.

\cite{Macci__2004} studied such models of coupled dark matter-dark energy using N-body simulations, with a simplified treatment of the coupling. It was found that, for strong coupling, the DM density profiles tended towards higher concentrations, exacerbating the cusp core problem. However, the study of \cite{Baldi_2009} found conflicting results for a similar coupling, showing a change in the slope of the density profile in the other direction, with a decrease in the central densities of the innermost regions of the DM halos.

A thorough study by \cite{Li_2011,Li_2011b} undertook a complete analysis of the consequences of these density-coupled DM-DE models. To begin with, the linear power spectrum was analysed where the contribution of baryons and DM was separated, observing that the presence of the coupling in the power spectrum at small scales shows an increase in the number of structures compared to $\Lambda$CDM, with this increase starting at a very early stage, even being relevant as early as $z=49$. Thus to be consistent, for these kinds of models, it is necessary to use initial conditions for the N-body simulations that differ from those of $\Lambda$CDM.

In addition it was found that the coupling effect leads to a modified non-linear matter power spectrum and mass function. As regards halo profiles it was shown that it is possible to see a reduction in the inner density profile, as compared to $\Lambda$CDM, although this suppression of the inner density is reduced for large couplings. In \cite{Li_2011b} the contributions of various effects in the coupled model were examined: the modified background expansion, varying particle mass, fifth force effects and finally the presence of a velocity dependent force. It was found that the first effect, the modified background expansion, is by far the most consequential for structure formation in these models.

 In \cite{Baldi_2012} the CODECs project is discussed, which significantly extended the explored  parameter space of such models, with both large-scale (L-CoDECS) and small-scale (H-CoDECS) models of dark matter density-coupled to dark energy, with the scalar field $\phi$ evolving according to a potential $V(\phi)$ of the exponential form. For all cases, they normalized all the models based on the same CMB amplitude, so for each simulation, they used different initial conditions. It was found that the coupling effects could break the degeneracy between DE and $\sigma_8$ at linear scales, given that the amplitude of the linear power spectrum exhibits a faster time evolution compared to $\Lambda$CDM. It was further found that for many coupled DM-DE models there is significant enhancement in structure formation in the non-linear regime leading to a modification of the halo mass function (HMF). While there is degeneracy, again, with $\sigma_8$, this can again be broken by the redshift dependence of the HMF.

In \cite{Simpson_2010} "dark scattering" models are discussed, which consider a momentum exchange between DM and DE. In this model, where the dark energy is treated as a fluid, a drag term in the velocity perturbation arises. In \cite{Bose_2018}, they analysed such models using various DE equations of state and various interaction cross-sections. They found that the effect of the interaction on linear perturbations acts efficiently to suppress/increase the amplitude of the power spectrum for large/small scales respectively.

In \cite{Baldi&simpson_2015}, they developed N-body simulations for the dark scattering models proposed by \cite{Simpson_2010}, using a dark energy fluid with equation of state parameter $w$, for $w > -1$ and $w < -1$. They found that the effect of DE-DM scattering on the linear power spectrum suppresses the power for $w > -1$ and increases it for $w < -1$. While in the nonlinear case, the effect is reversed, showing an increase for $w > -1$ and a suppression for $w < -1$ at z = 0. They further analysed the HMF, finding that the effect of scattering results in a significant increase (decrease) of the halo abundance over the whole mass range for $w > -1$ ($w < -1$). In addition, they analysed the velocity dispersion of the halos, finding that an increase in the scattering parameter leaves an increase in the velocity dispersion for all mass ranges when $w > -1$, while for $w < -1$, they did not find significant deviations.

In \cite{Baldi&simpson_2017} the same authors again performed N-body simulations, this time considering a time-evolving equation of state for the dark energy $w_{DE}$, the results of which were compared with \cite{Baldi&simpson_2015}, with the time-dependent equation of state leading to a weaker impact of the coupling at non-linear scales. Thus the amplification found in the power spectrum in the previous study is significantly suppressed in this case. These results suggest a possible avenue to reconcile low and high redshift observables.

The dark scattering model of \cite{Simpson_2010} has some resemblance to the \textit{Type-3} model proposed in \cite{Pourtsidou_2013}. This momentum transfer model considers the dark energy component as a scalar field, rather than a fluid, and leads to a significant modification to the equation of motion for the dark matter. Interestingly, the coupling is absent at the background level, unlike density coupled models. Furthermore, some of the modifications to the DM equation of motion are proportional to the DE density contrast, these latter modifications being absent in the dark scattering models. Given the presence of these additional terms the Type-3 model of \cite{Pourtsidou_2013} is not reducible to the dark scattering model of \cite{Simpson_2010}, as discussed in \cite{Skordis_2015}, and constitutes a new class of coupled DE-DM models.

This class of models was subsequently studied in \cite{Pourtsidou_2016,Chamings2020}, with a negative coupling constant, where it was found that the interaction suppressed structure growth, again possibly reconciling some of the tensions in CMB and LSS observations. These models have been further confronted with observations very recently in \cite{SpurioMancini2021}.

In this paper, we will focus on the study of the Type-3 model given in \cite{Pourtsidou_2013}. We will analyze, using N-body simulations, the impact of this coupling on the growth of structures as well as the influence (if any) on the shape of the power spectrum and halo properties by comparing with simulations of uncoupled models. We also briefly consider a $\Lambda$CDM model for reference. We are primarily interested in the implications for DM halos and the small-scale problems of the standard model discussed above.

\section{Theory and simulations}

We now discuss the theoretical background of the model we are considering and show how this is implemented in a cosmological N-body code. For full details of the theory we refer the reader to \cite{Pourtsidou_2013} and \cite{Skordis_2015}, which we closely follow in this section. Note that we work in units of $G = c = 1$.

\subsection{Equations of motion}

Dark matter is treated as a perfect fluid, in the usual manner, while the energy-momentum tensor of the scalar field for Type-3 models is written as
\begin{equation}
    T^{(\phi)}_{\mu \nu} = F_Y \phi_\mu \phi_\nu - F g_{\mu \nu} - Z F_Z u_\mu u_\nu.
\end{equation}
where
\begin{equation}
\begin{split}
Y &= \frac{1}{2}\phi_\mu \phi^\mu\\
Z &= u^\mu \phi_\mu,
\end{split}
\end{equation}
and $F = F(Y,Z,\phi)$ is some function. $F_Y$ and $F_Z$ denote derivatives of this function with respect to $Y$ and $Z$ respectively. The dark matter fluid 4-velocity is given by $u^{\mu}$ and $\phi_\mu \equiv \partial_\mu \phi$. The equations of motion for the scalar field and the dark matter fluid are given by
\begin{equation}
    \nabla_\mu (F_Y \phi^\mu + F_Z u^\mu) - F_\phi = 0,
\label{eom_phi}    
\end{equation}
and
\begin{equation}
    u^{\nu} \nabla_{\nu} \rho + \rho \nabla_{\nu} u^{\nu} = 0.
\label{eom_cdm}
\end{equation}
Note that the latter equation is just the standard equation of motion for an uncoupled pressureless perfect fluid. Thus the coupling has no direct effect on the DM density continuity equation. The momentum transfer equation is given by
\begin{equation}
(\rho - ZF_Z)u^{\beta}\nabla_{\beta}u_{\mu} = \nabla_{\beta}(F_Zu^{\beta})\tilde{\phi}_{\mu} + F_ZD_{\mu} Z,
\label{momentum_eq}
\end{equation}
where $D_{\mu} = q^{\nu}_{\mu} \nabla_{\nu}$ is the spatial derivative operator given in terms of the projection operator $q^\nu _\mu \equiv u_{\mu}u^{\nu} + \delta_{\mu}^{\nu}$, and $\tilde{\phi}_{\mu} = q^{\nu}_{\mu} \nabla_{\nu} \phi = D_{\mu} \phi = \partial_{\mu} \phi + u^{\nu}u_{\mu} \partial_{\nu} \phi$ is the spatial projection of the derivative of the scalar field. Note that this equation, in the absence of a coupling, is simply the standard geodesic equation for the dark matter fluid which reduces to the standard (pressureless) Euler equation in the Newtonian limit. Coupled quintessence is given by the choice $F = Y + V(\phi) + \gamma(Z)$, which we assume from now on.

To more easily connect with the Newtonian limit, as is relevant for our N-body simulations, we switch to the Newtonian gauge (in \citealp{Pourtsidou_2013} the synchronous gauge is used) described by the following line element:
\begin{equation}
    ds^2 = a^2(\tau)[-(1+2\Psi)d\tau^2 + (1+2\Phi)\delta_{ij}dx^i dx^j],
\label{ds_perturbed}    
\end{equation}
where $\Phi$ and $\Psi$ are spatial scalars and $\delta_{ij}$ is the 3-dimensional Kronecker delta (we always assume flat space), and the perturbed fluid 4-velocities (to linear order) in this gauge are,
\begin{eqnarray*}
    u_0 &=& -a(1 + \Psi), \\
    u_i &=& a v_i,
\end{eqnarray*}
where $v_i$ is the velocity perturbation of the fluid and $\Psi$ is one of the previously defined scalar components of the perturbed metric. 

The evolution of the cold dark matter fluid at the background level is given by the standard equation of motion in (\ref{eom_cdm}):
\begin{equation}
    \dot{\bar{\rho}} + 3\mathcal{H}\bar{\rho} = 0,
\end{equation}
and the evolution of the CDM fluid perturbations are
\begin{equation}
    \dot{\delta} + \theta + 3\dot{\Phi} = 0
\end{equation}
where the dot denotes a derivative with respect to conformal time, and we define $\theta \equiv \nabla_i v_i$. The CDM density, including first order perturbations, has been expressed as $\rho = \bar{\rho}(1+\delta)$. Again, we see that the density evolution is unaffected by the presence of the coupling. To simplify our notation we will from now on refer to the background CDM density with $\rho$.

Turning to the scalar field, the background evolution is given by equation~(\ref{eom_phi}) as
\begin{equation}
    \ddot{\phi} - \gamma_{ZZ}\ddot{\phi} + 2\mathcal{H}\dot{\phi} + \gamma_{ZZ}\mathcal{H}\dot{\phi} - 3a\mathcal{H}\gamma_Z + a^2V_\phi = 0.
\label{scalar_field_background}
\end{equation}

From equation (\ref{eom_phi}) at first order in the perturbations we obtain
\begin{equation}
\begin{split}
V_{\phi \phi} \varphi a^2 &+ 3\gamma_Z \Psi a \mathcal{H} + 3\gamma_Z a \dot{\Phi} - \gamma_Z a \nabla^2 \theta - \gamma_{ZZZ} \frac{\ddot{\bar{\phi}}}{a} \dot{\varphi}\\ 
&+ \gamma_{ZZZ} \frac{\dot{\bar{\phi}}}{a} \dot{\varphi} \mathcal{H} + 2\gamma_{ZZ}\Psi \ddot{\bar{\phi}} - 2\gamma_{ZZ}\Psi \dot{\bar{\phi}} \mathcal{H} + \gamma_{ZZ}\dot{\Psi}\dot{\bar{\phi}} \\
&- \gamma_{ZZ}\ddot{\varphi} - 2\gamma_{ZZ} \dot{\varphi} \mathcal{H} - 2\Psi \ddot{\bar{\phi}} - 4\Psi\dot{\bar{\phi}}\mathcal{H} - 3\dot{\Phi}\dot{\bar{\phi}} - \dot{\Psi}\dot{\bar{\phi}} + \ddot{\varphi} \\
&- \nabla^2 \varphi + 2\dot{\varphi} \mathcal{H} = 0
\label{scalar_field_perturbation}
\end{split}
\end{equation}

Since we want to take our analysis to small scales, it is useful to pass the equations to Fourier space, as is standard. Therefore each perturbed quantity $\chi$ and its derivatives can be substituted as follows,
\begin{eqnarray}
\chi(x, \tau) &\rightarrow \chi(\tau), \\
\nabla \chi(x, \tau) &\rightarrow k \chi(\tau), \\
\nabla^2 \chi(x, \tau) \equiv \nabla_i \nabla^i \chi(x, \tau) &\rightarrow k^2 \chi(\tau).
\end{eqnarray}

Applying this to equation (\ref{scalar_field_perturbation}) and taking the Newtonian limit\footnote{The requirement of non-relativistic velocities is implicit in the gauge choice, whereby the DM fluid velocity perturbation is assumed to satisfy $|v| \ll 1$.} of $k\gg\mathcal{H}$ (i.e. modes well within the horizon) the equation simplifies enormously to 
\begin{equation}
\label{eq:varphi}
\varphi = a\gamma_z\theta.
\end{equation}
We leave a more careful treatment of the scalar field perturbation, where this would be explicitly calculated using a numerical lattice field theory approach, to future work.

Using equation~(\ref{eq:varphi}) we can now write the momentum transfer equation (\ref{momentum_eq}) as
\begin{equation}
\begin{split}
\dot{\theta} + \mathcal{H}\theta + \Psi  = & \frac{1}{a\bar{\rho} - \gamma_Z \dot{\bar{\phi}}} \left[2\gamma_Z \dot{\bar{\phi}} \theta \mathcal{H} + 3a\gamma_Z^2 \theta \mathcal{H} - \gamma_Z \Psi \dot{\bar{\phi}} \right. \\
&\left. + \gamma_Z \ddot{\bar{\phi}} \theta + a \gamma_Z^2 \mathcal{H} \theta + a \gamma_Z \gamma_{ZZ} \dot{\bar{Z}} \theta + a \gamma_Z^2 \dot{\theta} \right] \\
- & \frac{1}{a^2 \bar{\rho} - a \gamma_Z \dot{\bar{\phi}}} \left[\gamma_{ZZ}\dot{\bar{\phi}}^2 \theta \mathcal{H} - \mathcal{H} \gamma_Z \gamma_{ZZ} \dot{\bar{\phi}} \theta \right. \\
&\left. + \gamma_{ZZ} \dot{\bar{\phi}} \ddot{\bar{\phi}} \theta + \gamma_Z \gamma_{ZZ} \dot{\bar{\phi}} \theta \right]
\end{split}
\label{momentum_transfer_eq_2}
\end{equation}
where $\dot{\bar{Z}} = 1/a(\ddot{\phi} + \mathcal{H}\dot{\phi})$. This is the modified Euler equation for momentum-coupled quintessence in a general form, where we have not yet selected the precise form of the coupling. In the absence of the coupling i.e. for $\gamma = 0$, the entire right hand side of equation~(\ref{momentum_transfer_eq_2}) is zero and we recover the standard Euler equation for the dark matter fluid. This equation must now be implemented in our N-body simulations.

\subsection{Modified Euler equation} \label{section_modified_euler_eq}

We follow \cite{Pourtsidou_2013} and define the coupling as
\begin{equation}
    \gamma(Z) = \gamma_0 Z^2
    \label{eq:gammaZ}
\end{equation}
where $\gamma_0$ is a constant whose value is assumed to be in the range $0 \leq \gamma_0 < 1/2$. Note that a negative value for $\gamma_0$ may in fact lead to more favourable observational consequences due to a reduction in structure at the linear level compared to the standard model, as discussed in \cite{Pourtsidou_2016}. As we will see, however, a large positive coupling can lead to reduced structure at non-linear scales. The equation (\ref{momentum_transfer_eq_2}) becomes
\begin{equation}
(1+h_1) \dot{v}_i  + (1+h_2) \mathcal{H}v_i + (1+h_3)\nabla_i \Psi = 0
\label{eq:modified_euler_eq}
\end{equation}
where the coefficients $h_1$, $h_2$ and $h_3$ are
\begin{equation}
\begin{split}
h_1 &= \frac{4\gamma_0^2 \dot{\phi}^2}{a^2\rho - 2\gamma_0\dot{\phi^2}}, \\
h_2 &= \frac{ (8\gamma_0^2 - 2\gamma_0)\dot{\phi}^2 + (8\gamma_0^2 - 4\gamma_0)\dot{\phi}\ddot{\phi}  \frac{1}{\mathcal{H}}}{ a^2\rho - 2\gamma_0\dot{\phi}^2},\\
h_3 &= \frac{ 2\gamma_0\dot{\phi}^2}{ a^2\rho - 2\gamma_0\dot{\phi}^2}.
\end{split}
\label{h_vals}
\end{equation}
In the $h_2$ term, we can replace $\ddot{\phi}$ using the evolution equation of the background field\footnote{Here we can see the presence of a strong coupling problem, as discussed in \cite{Pourtsidou_2013}, when $\gamma_0 = 1/2$. The largest value of $\gamma_0$ that we consider is $\gamma_0 = 0.3$.}, given by equation (\ref{scalar_field_background}), with 
\begin{equation}
\ddot{\phi}(1-2\gamma_0) + 2\mathcal{H}\dot{\phi}(1-2\gamma_0) + a^2 V_{\phi} = 0.
\label{scalar_field_background2},
\end{equation}
where we have used equation (\ref{eq:gammaZ}). Note that the coupling constant appears in this equation for the background evolution of the scalar field as an effective rescaling of $\phi$. Thus, $h_2$ can be written as \begin{equation}
h_2 = \frac{ 4\gamma_0 (\frac{3}{2} - 2\gamma_0)\dot{\phi}^2 + 4\gamma_0\dot{\phi} a^2 V_{\phi}/\mathcal{H}}{a^2\rho - 2\gamma_0\dot{\phi}^2}.
\end{equation}
From this, we can see that in the absence of the coupling we have $h_1 = h_2 = h_3 = 0$, and the Euler equation reduces to its standard form. In the presence of the coupling, however, we see that both the cosmological friction as well as the effective gravitational force acting on the DM are modified.

We can also now immediately see in what circumstance we would have modified dynamics, as compared with the uncoupled case. One might expect, given that the $h_i$ are all proportional to $\dot{\phi}^2$, that they would be negligible at late times. If the denominators in equation~(\ref{h_vals}) approach zero, however, then the values of $h_i$ will grow without bound. Due to the positivity of all quantities in both terms in the denominator we can therefore state that there will be a significant modification to the dynamics when
\begin{equation}
    a^2 \rho \approx 2\gamma_0 \dot{\phi}^2.
\end{equation}
Given that the dark matter density evolves as for the standard case, we can write the condition for large deviations from the standard dynamics as
\begin{equation}
    2a\gamma_0 \dot{\phi}^2 \approx \rho_0
\end{equation}
where $\rho_0$ is the present-day dark matter density. As we will see later, this condition is satisfied at late times for all of our models.

\subsection{Modification of the N-body solver}

For our cosmological N-body simulations we use the well-known RAMSES code \citep{Teyssier_2002}, which is a grid-based AMR code, using a particle-mesh (PM) scheme to evolve the dark matter particle distribution.

To write our equation in the code, we must first take into consideration the so-called supercomoving coordinates \citep{Martel_1998} used in RAMSES, which are defined as
\begin{equation}
\begin{split}
\vec{v} &= H_0 L \frac{1}{a}\tilde{\vec{u}}, \\
\vec{x} &= \frac{1}{a}\frac{\tilde{\vec{x}}}{L}, \\
dt &= a^2 \frac{d\tilde{t}}{H_0}, \\
\Psi &= \frac{L^2H_0^2}{a^2} \tilde{\Phi},
\end{split}
\label{coord_supercomoviles}
\end{equation}
where $L$ is the length of the simulation box. The coordinates denoted with a tilde are the supercomoving coordinates. To simplify the notation, we will apply the transformation and then remove the tildes. Thus, using equation (\ref{coord_supercomoviles}) in equation (\ref{eq:modified_euler_eq}) we get
\begin{equation}
    \frac{d\vec{u}}{dt} = -\frac{h_2 - h_1}{1 + h_1}a^2 \frac{H}{H_0} \vec{u} - \frac{1+h_3}{1+h_1} \vec{\nabla} \Phi.
    \label{euler_ec_standardform}
\end{equation}
Thus when $h_1 = h_2 = h_3 = 0$, we return to the standard form $\frac{d\vec{u}}{dt} = -\vec{\nabla}_{x} \Phi$, which is simply Newton's second law for a conservative force given by a potential $\Phi$. Transforming the Euler equation to supercomoving coordinates, in the uncoupled case, eliminates the cosmological friction term, simplifying the calculations in the code. In the presence of the coupling, however, the cosmological friction term is explicitly present, even in supercomoving coordinates. Note that equation~(\ref{euler_ec_standardform}) uses the Hubble parameter with respect to physical time.

We now have the modified Euler equation (\ref{euler_ec_standardform}) in a form in which it may be discretised and solved numerically. In RAMSES, a finite difference approximation is used to resolve the equations of motion, using a Leapfrog scheme. Given an acceleration $-\nabla \phi^n$ at a time $t^n$, with particle positions $x^{n}_p$ and velocities $v^{n}_p$, the velocities are updated by a half timestep using the potential at $t^n$ and then the positions are updated using these updated velocities, according to
\begin{equation}
\begin{split}
    v^{n+1/2}_p &= v^{n}_p - \nabla \phi^n \Delta t^n/2, \\
    x^{n+1}_p &= x^{n}_p + v^{n+1/2}_p \Delta t^n/2,
\end{split}
\end{equation}
which is then followed by a full update of the velocity using the updated gravitational potential:
\begin{equation}
    v^{n+1}_p = v^{n+1/2}_p - \nabla \phi^{n+1} \Delta t^n/2.
\end{equation}
Note that the time-step in RAMSES is adaptive, thus we write $\Delta t^n$. To connect with the implementation of the modified Euler equation in the code, we write the finite difference update of the velocity as
\begin{equation}
    \frac{v^{n+1/2}_p - v^{n}_p}{(1/2)\Delta t^n} = F,
\end{equation}
where $F$ is the force acting on the particle.
This is simply a finite difference approximation to the differential equation (\ref{euler_ec_standardform}) in the absence of coupling. Thus, we can easily modify the velocity update as required to implement equation (\ref{euler_ec_standardform}) in the following way:
\begin{equation}
    v^{n+1/2}_p = v^n_p - \frac{h_2 - h_1}{1+h_1}a^2\frac{H}{H_0}v^n_p \Delta t^n/2 + \frac{1+h_3}{1+h_1}F \Delta t^n/2.
\label{vel_timestep_hvalues}    
\end{equation}
We now define two new coefficients $\epsilon_1$ and $\epsilon_2$ to simplify the expression,
\begin{equation}
\begin{split}
\epsilon_1 &= 1 - \frac{h_2-h_1}{1+h_1}a^2\frac{H}{H_0}\Delta t^n/2 \\
\epsilon_2 &= \frac{1+h_3}{1+h_1}
\end{split}
\end{equation}
so finally equation (\ref{vel_timestep_hvalues}) becomes
\begin{equation}
        v^{n+1}_p = \epsilon_1 v^n_p + \epsilon_2 F \Delta t^n/2.
\label{eq_ramses}        
\end{equation}
This is the equation we have implemented in RAMSES. The standard dynamics is recovered by setting $\epsilon_1 = \epsilon_2 = 1$ which is equivalent to having all the $h_i$ equal to zero.

\subsection{Obtaining the background values}
Going back to the modified Euler equation (\ref{eq:modified_euler_eq}), we can see that the $h_i$ values (or, equivalently, the $\epsilon_1$ and $\epsilon_2$ coefficients in RAMSES) depend on background quantities such as $\rho$, $\phi$, and $\mathcal{H}$. To solve the evolution of these values we used a modified version of the CLASS code \citep{lesgourgues2011} which calculates the evolution of linear cosmological perturbations. CLASS includes the option to add a quintessence field to the matter-energy components of the Universe. Our modifications of the code were to include the coupling term in the Klein-Gordon equation (\ref{scalar_field_background}), the scalar field perturbation equation (\ref{scalar_field_perturbation}) and the momentum transfer equation (\ref{momentum_transfer_eq_2}), although we have used only the modified background Klein-Gordon equation (thus including the field rescaling) to calculate the evolution of the background quantities. We will, however, use the full perturbation equations momentarily to confirm that, for our specific models, there is a minimal impact on the CMB power spectrum. The scalar field potential used in our study is that of \cite{Albrecht_2000}, given by
\begin{equation}
V(\phi) = ((\phi - \beta)^{\alpha} + \Gamma)e^{-\lambda \phi}
\end{equation}
which is already included in CLASS.


\begin{table*}
\begin{tabular}{ c | c c c c c c c}
 \hline 
Model & Potential & $\Gamma$  & $\beta$ & $\lambda$ & $\alpha$ & $\phi$ & $\dot{\phi}$ \\
 \hline \hline
 A & $\Gamma e^{-\lambda \phi}$ & 1.0 & - & 1.597723e-1 & - & 100 & 10 \\
B & $[(\phi - \beta)^{\alpha} + \Gamma]e^{-\lambda \phi}$ & 0.001 & 34.8 & 2.432815e-1 & 2.0 & 100 & 10 \\
C & $[(\phi - \beta)^{\alpha} + \Gamma]e^{-\lambda \phi}$ & 20.0 & 3.8 & 9.347720e-1 & 17.0 & 100 & 10 \\
 \hline
\end{tabular}
\caption{Parameters of the scalar field potential used in our models. The scalar field values are in units of the reduced Planck mass $m_{P} = \sqrt{8 \pi G}$, and the dot indicates a derivative with respect to conformal time. For all models we consider three values of the coupling: $\gamma_0 = 0$, $0.15$ and $0.3$.}
\label{tabla_simulaciones}
\end{table*}

The parameter values for the potentials we have considered are given in Table \ref{tabla_simulaciones}. The CLASS code adjusts one parameter, chosen by the user, in order to satisfy the closure condition of the density parameters: $\sum_i \Omega_i = 1$. In our case this parameter was chosen to be $\lambda$. Note that the values of the other parameters have been based on those used in \cite{Albrecht_2000}, although we have considered rather large initial conditions for the scalar field and its derivative. The resulting evolution is not sensitive to the latter.

CLASS calculates the background evolution equations for all components of the Universe (including baryons and radiation), giving us tabulated values for all background quantities at a large number of redshifts. Obtaining these values, we then transform the conformal time given in the table calculated by CLASS to the superconformal time (see equation~\ref{coord_supercomoviles}) used in RAMSES.

For each model we obtain the background evolution (shown in \S  \ref{sec:background_Evolution}). We have added a small routine to RAMSES to read the table of background values generated by CLASS in order to calculate the values of the $\epsilon_1$ and $\epsilon_2$ coefficients in the modified Euler equation. The values are determined at the appropriate redshift by linear interpolation of neighbouring values in the table.

\subsection{Initial conditions and parameters for the N-body simulation}
The initial conditions for the simulations were generated using MUSIC \citep{Hahn_2011}, with a transfer function generated from CAMB \citep{Lewis_2011}, assuming a standard $\Lambda$CDM cosmology, with cosmological parameters given in Table~\ref{tabla_music}. We used CAMB for expediency given that the output files are compatible with MUSIC.

\begin{table}
\begin{center}
\begin{tabular}{ c | c }
 \hline
 Parameter & Value \\
 \hline \hline
 $H_0$ & 70 km s$^{-1}$ Mpc$^{-1}$\\
 $\Omega_{m}$ & 0.3 \\
 $\Omega_{\Lambda}$ & 0.7 \\
 $\Omega_b$ & 0.04 \\
 $\sigma_8$ & 0.88 \\
 $n_s$ & 0.96 \\
 \hline
 \end{tabular}
 \caption{The cosmological parameters used for generation of the initial conditions.}
 \label{tabla_music}
 \end{center}
 \end{table}

Given that the background evolution of the dark matter fluid is unaffected by the coupling in our models, the transfer function at high redshift is effectively identical to the uncoupled case and very similar to that of $\Lambda$CDM. In addition, leaving the transfer function fixed across all models allows us to generate identical initial conditions, thus simplifying the process of comparing the low-redshift results. We leave for future work the use of fully consistent simulations with appropriately modified initial conditions. 

The physical box size used for most of our simulations is 32 Mpc $h^{-1}$, the number of particles $N_{p} = 128^3$ and the initial redshift is $z_{ini} = 50$. We have also run some simulations of model C (with $\gamma_0 = 0$ and $\gamma_0 = 0.3$) in larger boxes of sizes $128$ Mpc $h^{-1}$ and $512$ Mpc $h^{-1}$ in order to study the power spectra at larger scales. These simulations use $256^3$ particles, and thus have limited mass resolution and are not used for the halo analysis. Throughout this text when we refer to ``large scales'' we are referring to scales of order the size of the simulation box. The simulation parameters are summarised in Table~\ref{table:technical_prperties}.

\begin{table}
\begin{center}
\begin{tabular}{c|ccc}
\hline
$N_{p}$ & $M_{p}$ [$M_{\odot}$ h$^{-1}$] & $L$ [Mpc h$^{-1}$] & $\Delta_x$ [kpc h$^{-1}$] \\
\hline \hline
$128^3$ & $\sim 1.3 \times 10^9$ & $32$ & $1.95$ \\
$256^3$ & $\sim 1.0 \times 10^{10}$ & $128$ & $7.8$ \\
$256^3$ & $\sim 6.7 \times 10^{11}$ & $512$ & $31.2$ \\
\hline
\end{tabular}
\caption{Technical properties of all our simulations. $\Delta_x$ refers to the maximum spatial resolution. \label{table:technical_prperties}}
\end{center}
\end{table}

All the models studied are summarized in Table \ref{tabla_simulaciones} with three additional cases given in Table \ref{tabla_simulaciones_coeffs}, which consider the individual contributions of each Euler coefficient as they are implemented in RAMSES: $\epsilon_1$ and $\epsilon_2$.
\begin{table}
\begin{center}
\begin{tabular}{ c | c c }
 \hline
Model & $\epsilon_1$ & $\epsilon_2$ \\
 \hline \hline
C* & 1 & 1 \\ 
C*1 & $1 - \frac{h_2-h_1}{1+h_1}a^2\frac{H}{H_0}\Delta t^n/2$ & 1 \\
C*2 & 1 & $\frac{1+h_3}{1+h_1}$\\
 \hline
\end{tabular}
\caption{
The two additional simulations C*1 and C*2 consider each coefficient in the Euler equation separately, using the background evolution of model C for $\gamma_0 = 0.3$. The model C* has the background evolution of model C with $\gamma_0 = 0.3$ and $\epsilon_1 = \epsilon_2 = 1$.} 
\label{tabla_simulaciones_coeffs}
\end{center}
\end{table}
The parameters for the scalar field potential have been chosen to produce a background evolution that is similar to that of $\Lambda$CDM. Our idea is not to deviate excessively from the background evolution of the standard model.

For the coupling parameter, we choose the uncoupled case $\gamma_0 = 0$, an intermediate coupling $\gamma_0 = 0.15$ and a strongly coupled model $\gamma_0 = 0.3$. Note that the $\gamma_0 = 1/2$ case is theoretically excluded \citep{Pourtsidou_2013}. It is also worth pointing out that the uncoupled case is not equivalent to $\Lambda$CDM, due to the presence of the (uncoupled) quintessence field, rather than a cosmological constant.

\section{Results}
We now present the results obtained from our simulations using the CLASS and RAMSES codes.

\subsection{Background evolution}\label{sec:background_Evolution}

\begin{figure}
  \centerline{\includegraphics[scale=0.25]{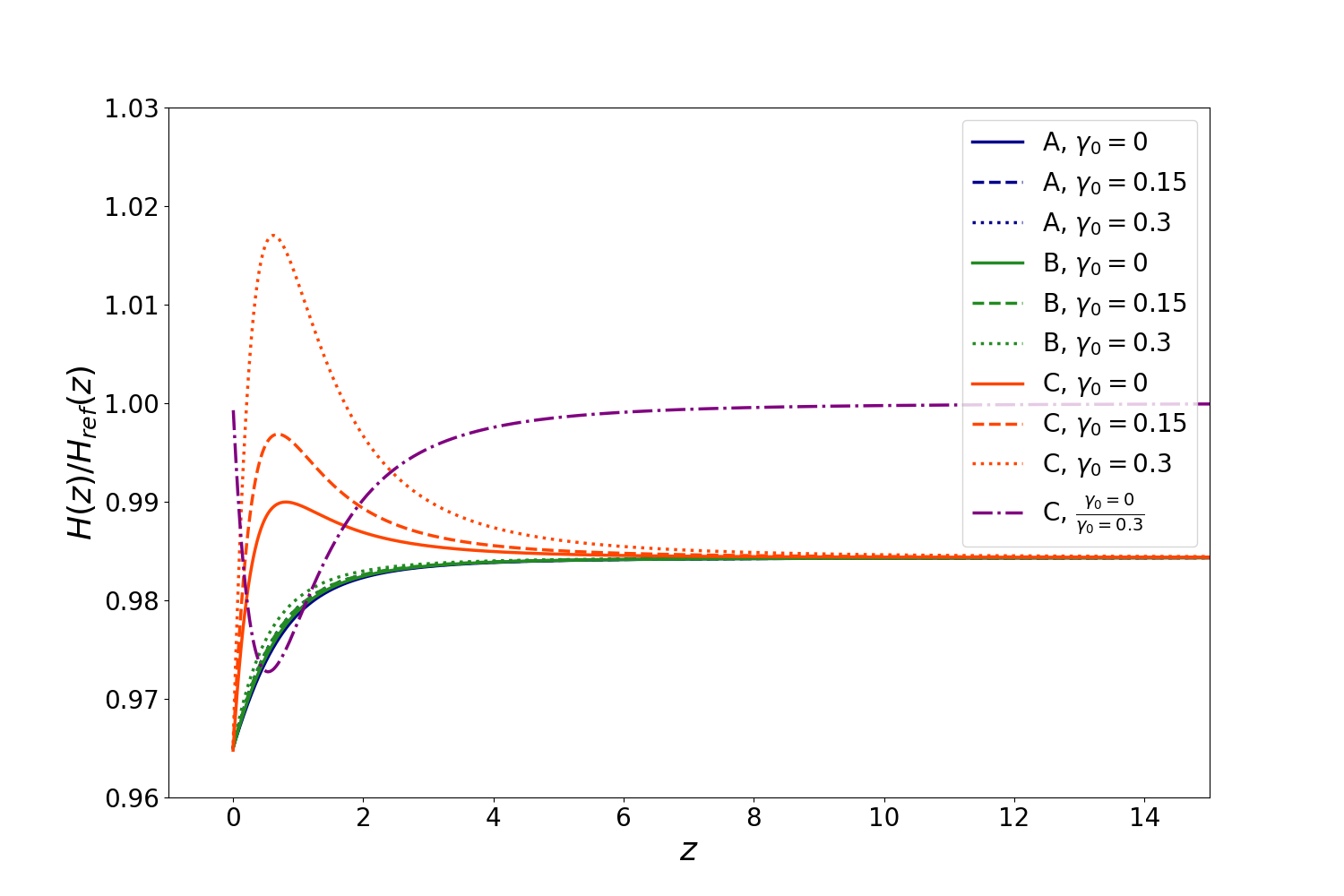}}
  \caption{Evolution of $H(z)/H_{\text{ref}}(z)$ for all models summarised in Table \ref{tabla_simulaciones}\label{omega_phi_evol}. In almost all cases $H_{\text{ref}} = H_{\text{LCDM}}(z)$. The exception is the purple dot-dashed line, which is for $H(z)$ of model C without coupling normalised by $H(z)$ of model C with $\gamma_0 = 0.3$.}
\end{figure}

From CLASS we can obtain the evolution of the density parameters $\Omega_i$ in our models, with $i$ corresponding to DM, DE, baryons and radiation. Their evolution is broadly consistent with that of $\Lambda$CDM. In Fig.~\ref{omega_phi_evol} we show the evolution of $H(z)$ normalised by the Hubble parameter of $\Lambda$CDM, as well as the values of $H(z)$ for model C with $\gamma_0 = 0$ normalised by $H(z)$ for model C with $\gamma_0 = 0.3$ to show the effect of the coupling on the background (due to the presence of $\gamma_0$ in equation ~\ref{scalar_field_background2}). As we can see in the figure, the values of $H(z)$ for our quintessence models at high redshift are a constant $\sim 1\%$ lower than those of $\Lambda$CDM. At low redshift, however, models A (blue lines) and B (green lines) deviate to a maximum of $\sim 3\%$ lower values of the Hubble parameter by $z=0$, independent of the value of the coupling constant. For model C, we see a more complex behaviour, where the Hubble parameter increases towards the $\Lambda$CDM value, before then decreasing again to very similar final values as seen for models A and B. In the case of model C with $\gamma_0 = 0.3$ (orange dotted line) $H(z)$ peaks at a value that slightly exceeds the standard cosmology, before dropping rapidly. For model C with $\gamma_0$ normalised by the Hubble parameter for the same model but with coupling $\gamma_0 = 0.3$ (purple dot-dashed line) we see that for large redshift the background evolution is identical in the two models, but at small redshifts the Hubble parameter for the coupled model takes smaller values than that of the uncoupled model. The consequences of this for the power spectra will be explored in Section~\ref{subsec:ps}.


\begin{figure}
  \centerline{\includegraphics[scale=0.25]{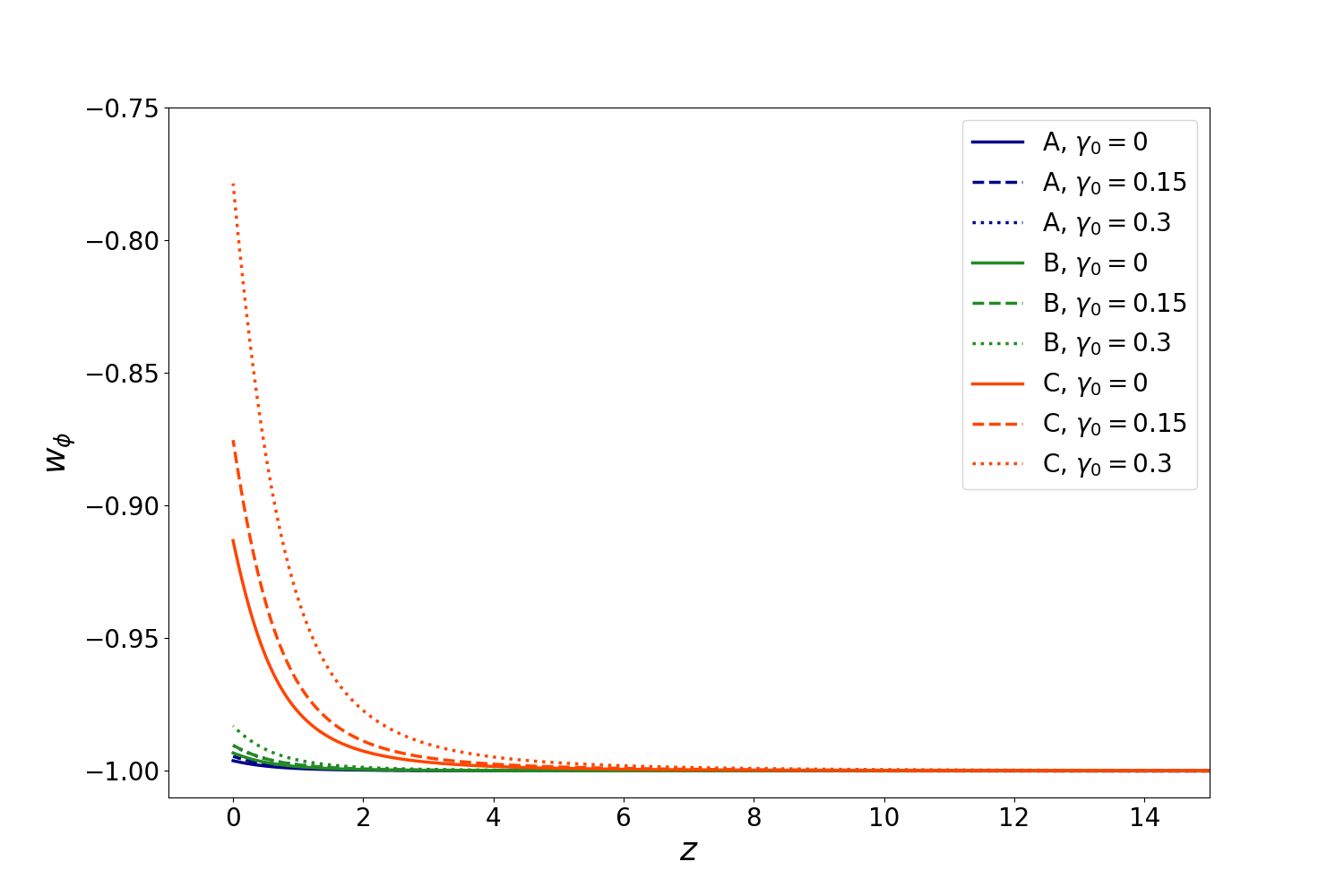}}
  \caption{The equation of state parameter $w_\phi$ for models A (blue lines), B (green lines) and C (orange lines).}\label{eos_models}
\end{figure}

The equation of state parameter $w_\phi$ for a quintessence model is given by
\begin{equation}
    w_\phi \equiv \frac{p_{\phi}}{\rho_{\phi}} = \frac{\dot{\phi}^2/2 - V(\phi)}{\dot{\phi}^2/2 + V(\phi)}.
\label{eos_quintessence}    
\end{equation}
and is shown in Fig.~\ref{eos_models}. For all models, $w_\phi$ begins with a value equal to $1$, which then decays rapidly to values close to $-1$. This transition occurs at very high redshift, well before the starting redshift of our N-body simulations, and occurs because of the form of the chosen potential. If we observe the evolution for each value of $\gamma_0$, we see an almost identical behaviour, with separation of the models as we approach z = 0. The values of $w_\phi(z=0)$ are given in Table \ref{tabla_eos}. The evolution is given by a dynamical equation of state whose final value $w(z = 0)$, for all our models, is $w > -1$. 

\begin{table}
\begin{center}
\begin{tabular}{ l | c c c}
 \hline
$w_\phi(z=0)$ & A  & B  & C \\
 \hline \hline
 $\gamma_0$ = 0 & -0.996 & -0.993 & -0.913 \\
 $\gamma_0$ = 0.15 & -0.994 & -0.990 & -0.875 \\
 $\gamma_0$ = 0.3 & -0.990 & -0.983 & -0.779 \\
 \hline
 \end{tabular}
 \caption{Equation of state $w_\phi(z=0)$ for our models A, B and C with different values of $\gamma_0$.\label{tabla_eos}}
 \end{center}
 \end{table}

\begin{figure}
  \centerline{\includegraphics[scale=0.25]{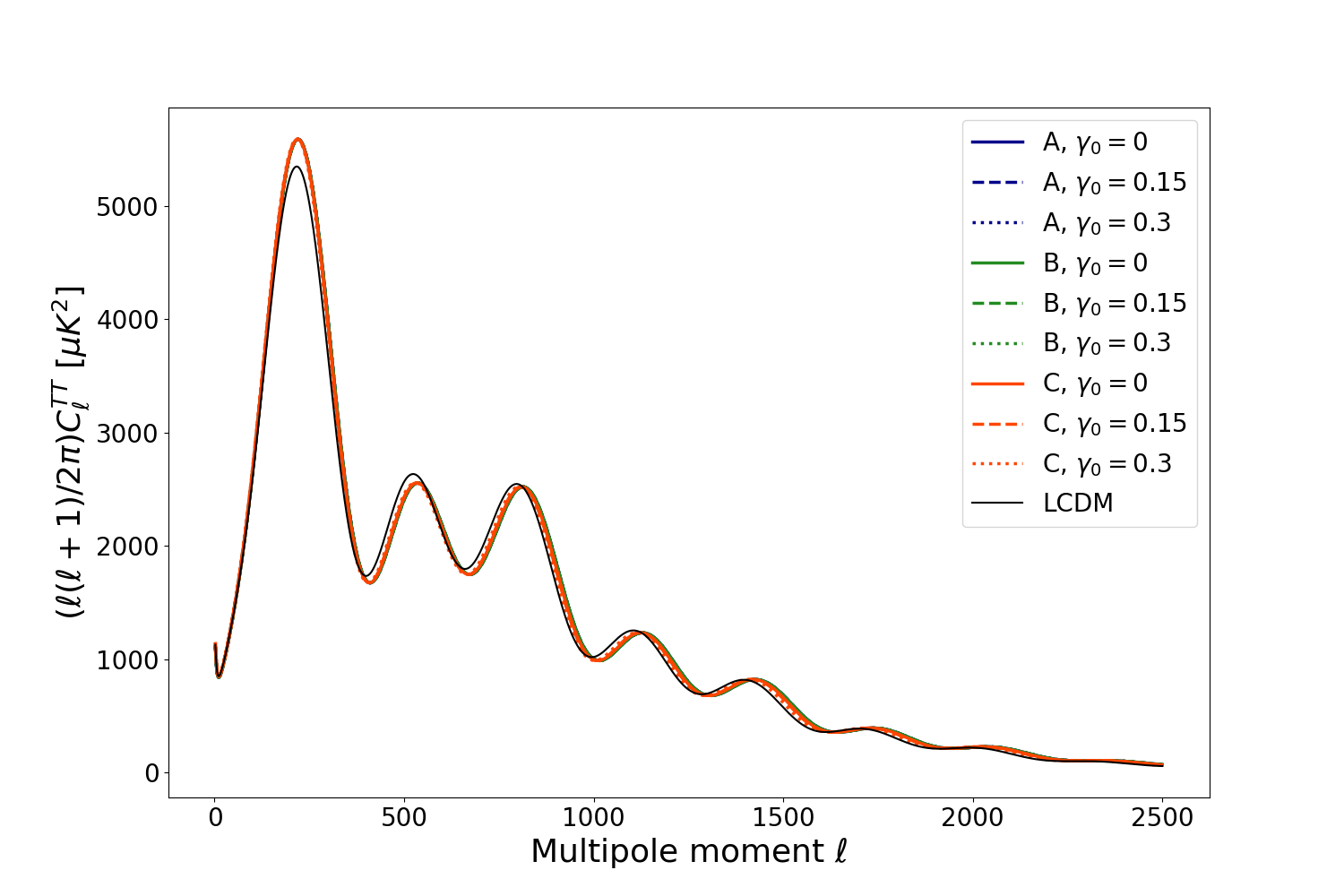}}
  \caption{The angular power spectrum of the CMB temperature fluctuations for LCDM (black line) and our models (coloured lines).\label{cmb_spectra}}
\end{figure}

To check that our models are consistent with CMB observations, we now compare the CMB temperature fluctuation power spectrum of our fiducial $\Lambda$CDM model with our coupled quintessence models in Fig.~\ref{cmb_spectra}. As we can see, there are only very minor deviations in the peaks of the power spectra, when comparing with $\Lambda$CDM, due to the slightly modified background evolution. It is worth noting, however, that the coupled quintessence models that we consider lead to CMB power spectra that are essentially identical, regardless of the potential or the coupling. 

In order to quantitatively understand the deviation in our modified Euler equation from the uncoupled case, we focus on equation (\ref{eq:modified_euler_eq}). From this equation we can directly estimate the magnitude of our modifications and how these might affect the movement of the particles. Dividing equation (\ref{eq:modified_euler_eq}) throughout by the coefficient of the acceleration term, we can refer to the coefficient of the cosmological friction term as $c_1$ and the coefficient of the gravitational force term as $c_2$, that is:
\begin{equation}
\begin{split}
c_1 &= \frac{1 + h_2}{1 + h_1} \\
c_2 &= \frac{1 + h_3}{1 + h_1}.
\end{split}
\end{equation}

\begin{figure}
  \centerline{\includegraphics[scale=0.25]{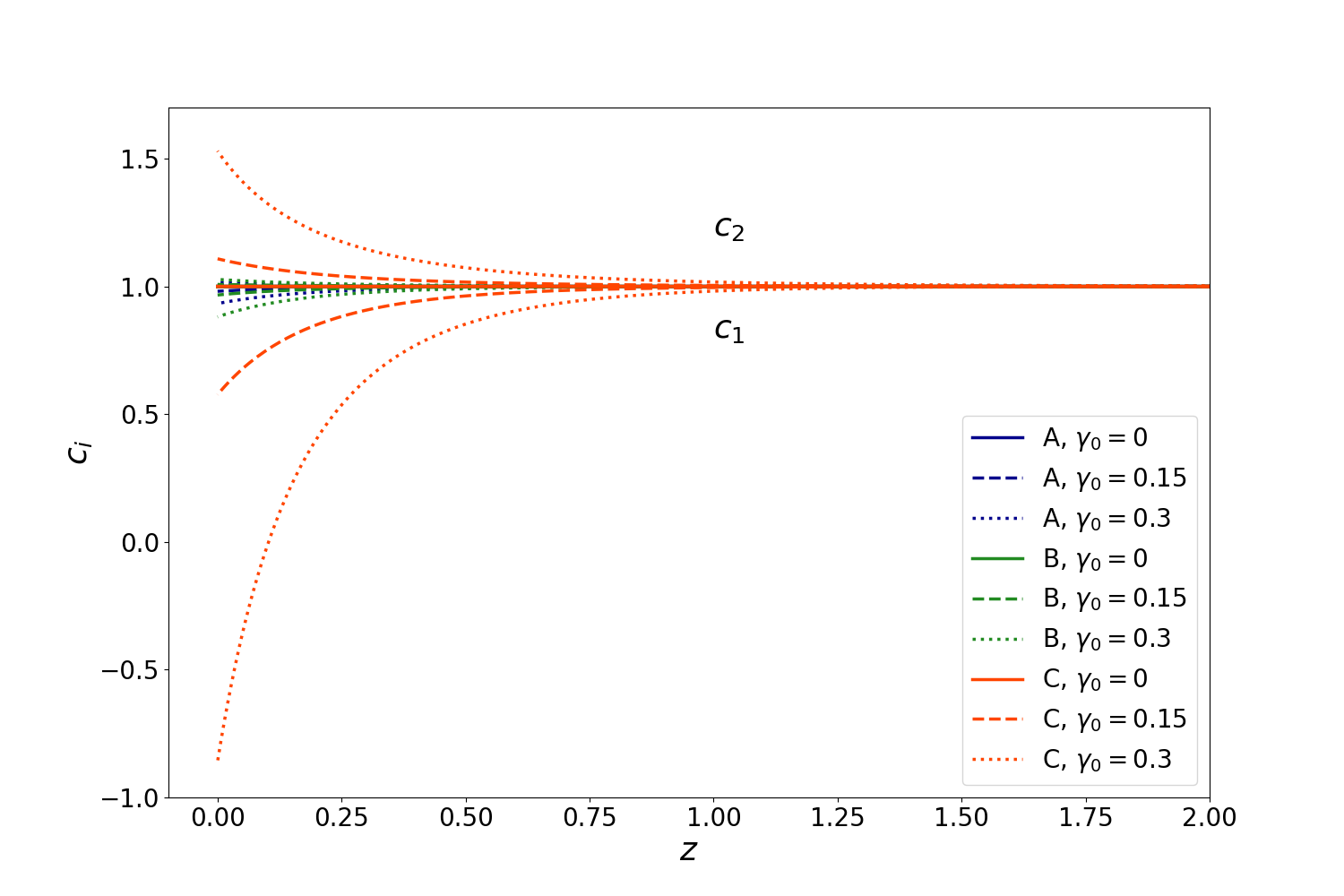}}
  \caption{Variation of the cosmological friction ($c_1$) and gravitational force ($c_2$) coefficients.}{\label{fig:euler_coeffs}}
\end{figure}

Fig.~\ref{fig:euler_coeffs} shows the evolution of these coefficients for all our models, which we compare with the uncoupled case (i.e., with $c_1$ = $c_2$ = 1). We plot models A (blue lines), B (green lines) and C (orange lines), distinguishing for each value of $\gamma_0$. We summarize the deviations of our models from the standard case in Table \ref{tabla_euler_coeff}. Considered as a percentage deviation from the standard case, we can see that the modified cosmological friction term always dominates over that of the modified gravitational force term, at least for the models considered in this study. This would suggest that, in the limit of weak coupling, all of our models would correspond to the dark scattering case. The cosmological friction and effective gravitational force in our models remains the same as the uncoupled case until z $\sim$ 2, indicating that the presence of the coupling only becomes relevant at low redshift.

As we approach z = 0, we see a reduction in the cosmological friction, which is particularly pronounced for $\gamma_0$ = 0.3 in model C (orange dotted line) with a change to negative values and a deviation of over $180\%$. It is well known that the cosmological friction term in the standard model acts to slow down the formation of structure, given that it is a force directed anti-parallel to the particle velocities. A reduction in the coefficient of this term thus implies a reduction in the effectiveness of the cosmological friction, meaning the particles will be less decelerated by the cosmological expansion as compared to the standard case. If this coefficient is equal to zero the cosmological friction is entirely cancelled out leading to unconstrained growth of the gravitational instability. This situation would only arise for a brief period of time in our models, however, due to the time-variation of the coefficients. In the extreme case of a negative coefficient the frictional force then acts parallel to the particle velocities and thus the cosmological friction term in this case becomes a kind of forcing (we will refer to this as the ``cosmological push'' throughout the rest of the paper). For model A we see that the deviation is much smaller for all $\gamma_0$ values, being $7\%$ for $\gamma_0$ = 0.3, while for model B the deviation reaches $12\%$ at the present time for $\gamma_0$ = 0.3. 

As for the evolution of $c_2$, we see that the general behavior is an increase in the coefficient of the gravitational force term. The variation for A reaches $0.4\%$ for $\gamma_0$ = 0.15 and 1.5\% for $\gamma_0$ = 0.3, while for model B we see that the gravitational force increases by $0.8\%$ and $2.7\%$ for $\gamma_0$ = 0.15 and 0.3, respectively. For model C, we see that the deviation rises significantly, reaching $53\%$ for $\gamma_0$ = 0.3. 

We will see later that both the modified gravitational force term and the modified cosmological friction term have an impact in modifying the evolution of structure, particularly in model C with the largest coupling.

It is worth pointing out that, in the extreme case of model C where $c_1 < 0$, due to the combined effects of the enhanced effective gravitational force and the cosmological push, we would expect to see increasingly ``hot'' gravitationally bound systems, i.e. the velocity dispersions of bound halos will increase with increasing $c_1$ and $c_2$, presumably leading to some kind of instability in the absence of a mechanism to stop the increase of these coefficients. In this paper we will not address the details of this instability or the virialisation process of the halos. We leave this for future work.

\begin{table}
\begin{center}
\begin{tabular}{ l | c c }
 \hline
Model & $c_1$ (z = 0) & $c_2$ (z = 0) \\
 \hline \hline
 A, $\gamma_0$ = 0.15 & 0.981 & 1.004 \\
 A, $\gamma_0$ = 0.3 & 0.933 & 1.015 \\
 B, $\gamma_0$ = 0.15 & 0.966 & 1.008 \\
 B, $\gamma_0$ = 0.3 & 0.881 & 1.027 \\
 C, $\gamma_0$ = 0.15 & 0.576 & 1.108 \\
 C, $\gamma_0$ = 0.3 & -0.859 & 1.530 \\
 \hline
\end{tabular}
\caption{Values of the coefficients of the modified Euler equation at z = 0.}
\label{tabla_euler_coeff}
\end{center}
\end{table}
 

\subsection{Density distribution and power spectrum}
\label{subsec:ps}
\begin{figure*}
  \centerline{\includegraphics[scale=0.5]{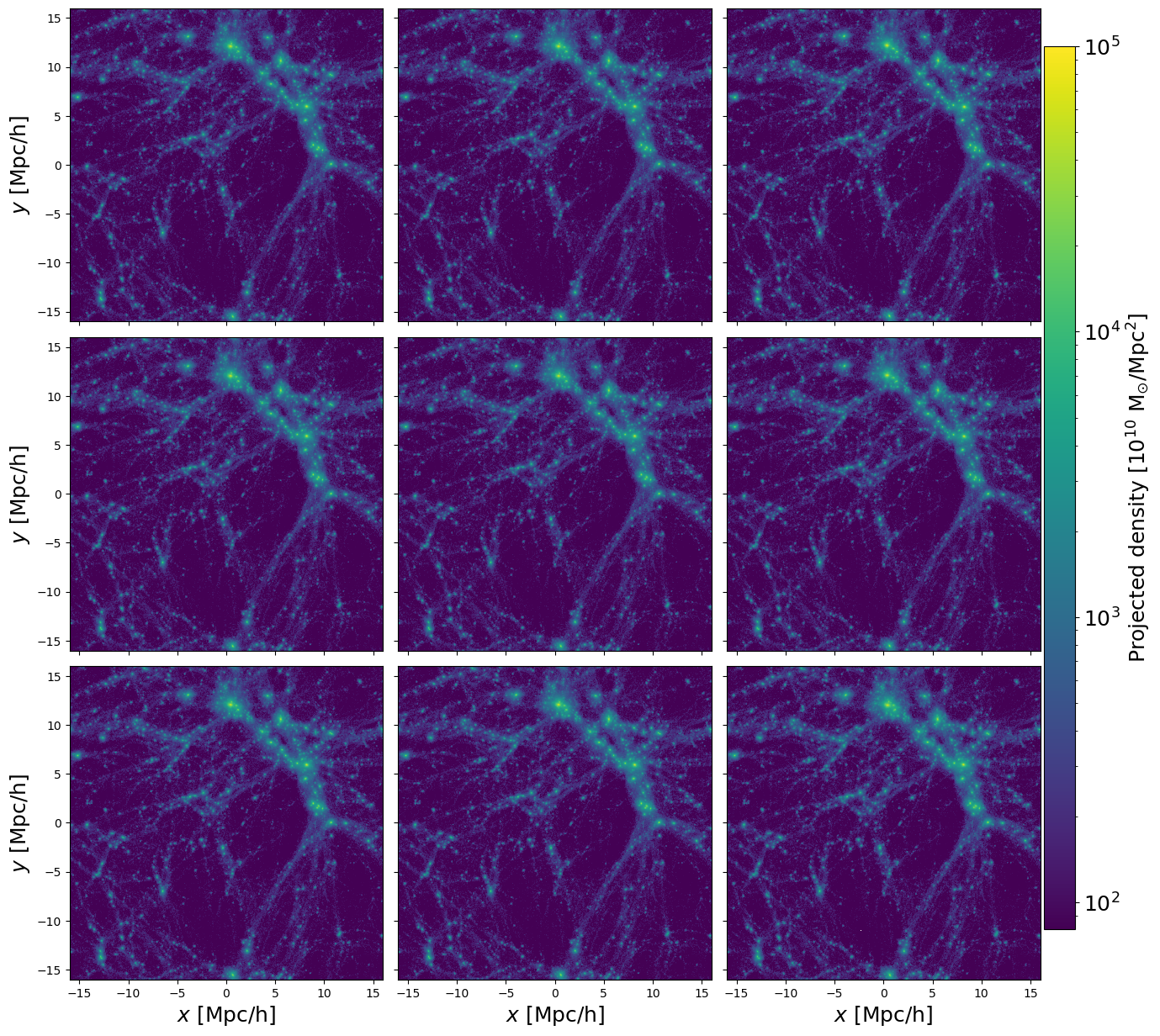}}
  \caption{Final projected particle distribution for models A (left column), B (middle column) and C (right column) at $z = 0$ (box size $32$ Mpc $h^{-1}$, with $128^3$ particles). The coupling is $\gamma_0$ = 0, 0.15 and 0.3 in the top, middle and bottom rows.\label{projected_particles}.}
\end{figure*}

\begin{figure*}
  \centerline{\includegraphics[scale=0.5]{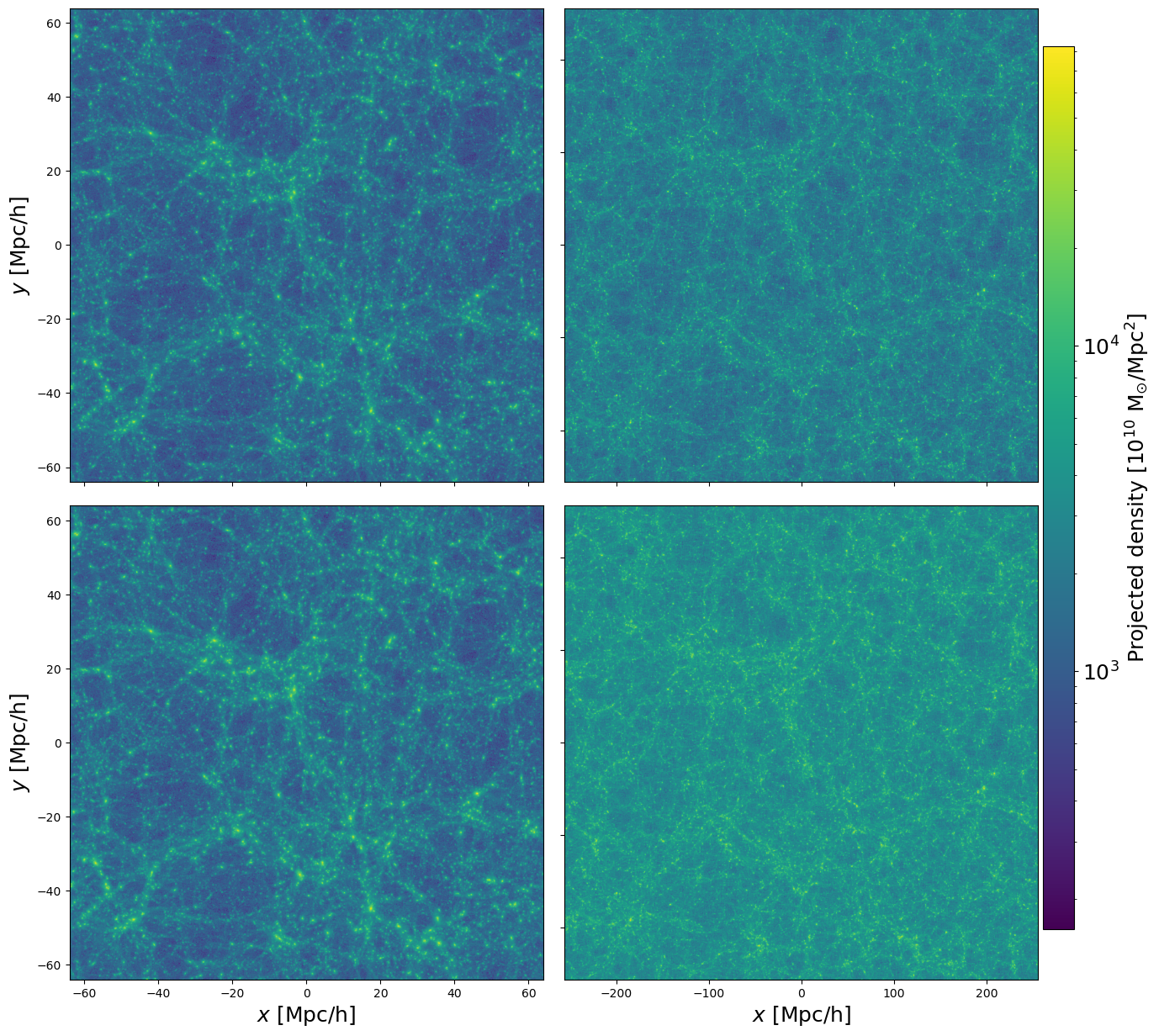}}
  \caption{Final projected particle distribution for model C with $\gamma_0 = 0$ (top row) and $\gamma_0 = 0.3$ (bottom row) at $z = 0$. \textit{Left column}: box size $128$ Mpc $h^{-1}$, \textit{right column}: box size $512$ Mpc $h^{-1}$. In all cases the particle number is $256^3$.\label{projected_particles_bigBox}}
\end{figure*}

We now turn to the results of our N-body simulations. For all of our results we will consider the final simulation snapshot at $z=0$. The projected particle density distributions for models A, B and C are shown in Fig.~\ref{projected_particles} for the small box of $32$ Mpc $h^{-1}$ and in Fig.~\ref{projected_particles_bigBox} for the larger boxes of $128$ Mpc $h^{-1}$ and $512$ Mpc $h^{-1}$. The density distribution is visually very similar across all models in Fig.~\ref{projected_particles} given that we have used identical initial conditions for all runs. Similarly, the structure produced by $z=0$ in the large box simulations of model C, as shown in Fig.~\ref{projected_particles_bigBox}, is visually very similar, regardless of whether the coupling is present.

\begin{figure}

  \centerline{\includegraphics[scale=0.25]{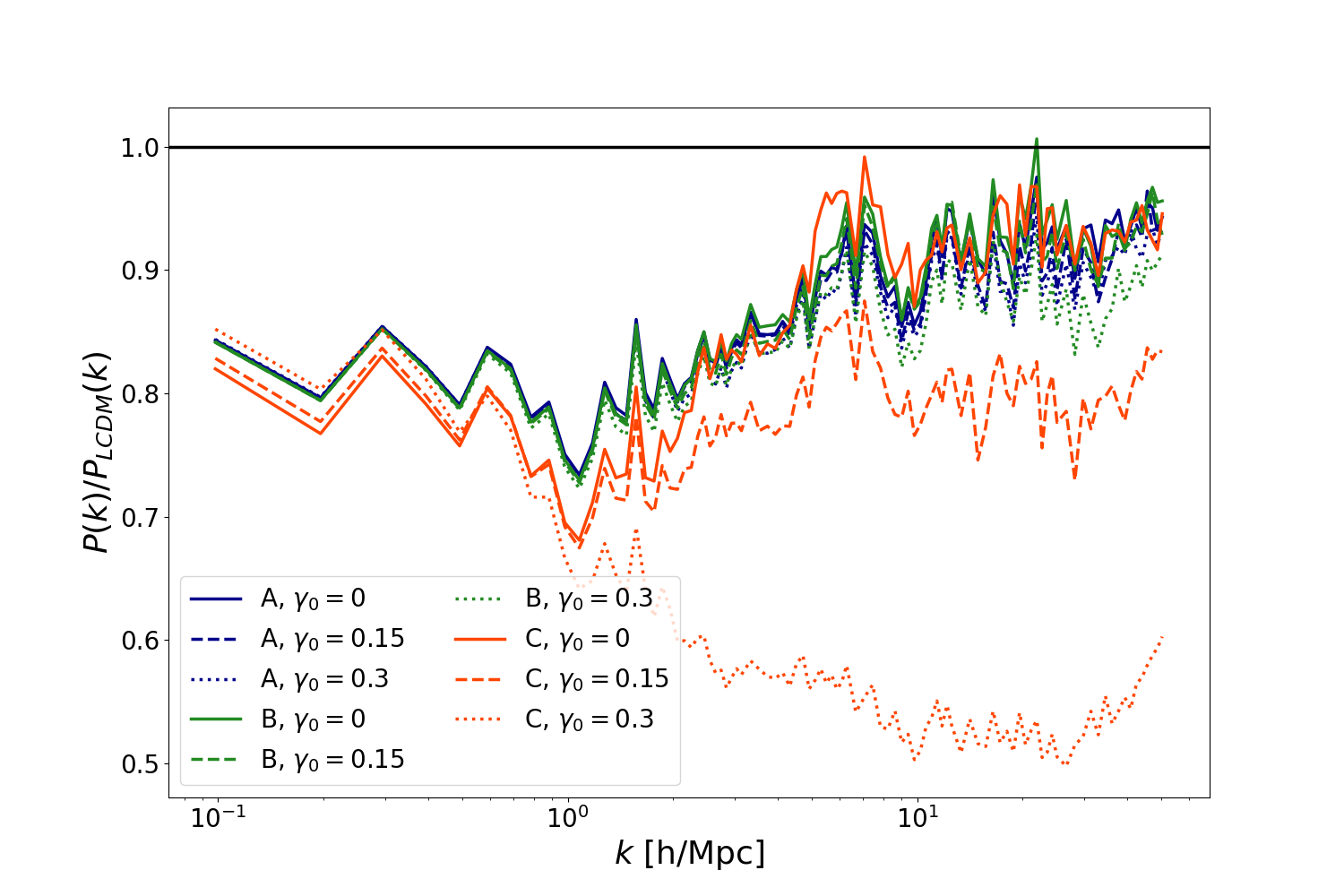}}
  \caption{Ratio of the power spectrum of models A (blue lines), B (green lines) and C (orange lines) with respect to $\Lambda$CDM. \label{ps_models}}
  
\end{figure}

\begin{figure}

  \centerline{\includegraphics[scale=0.25]{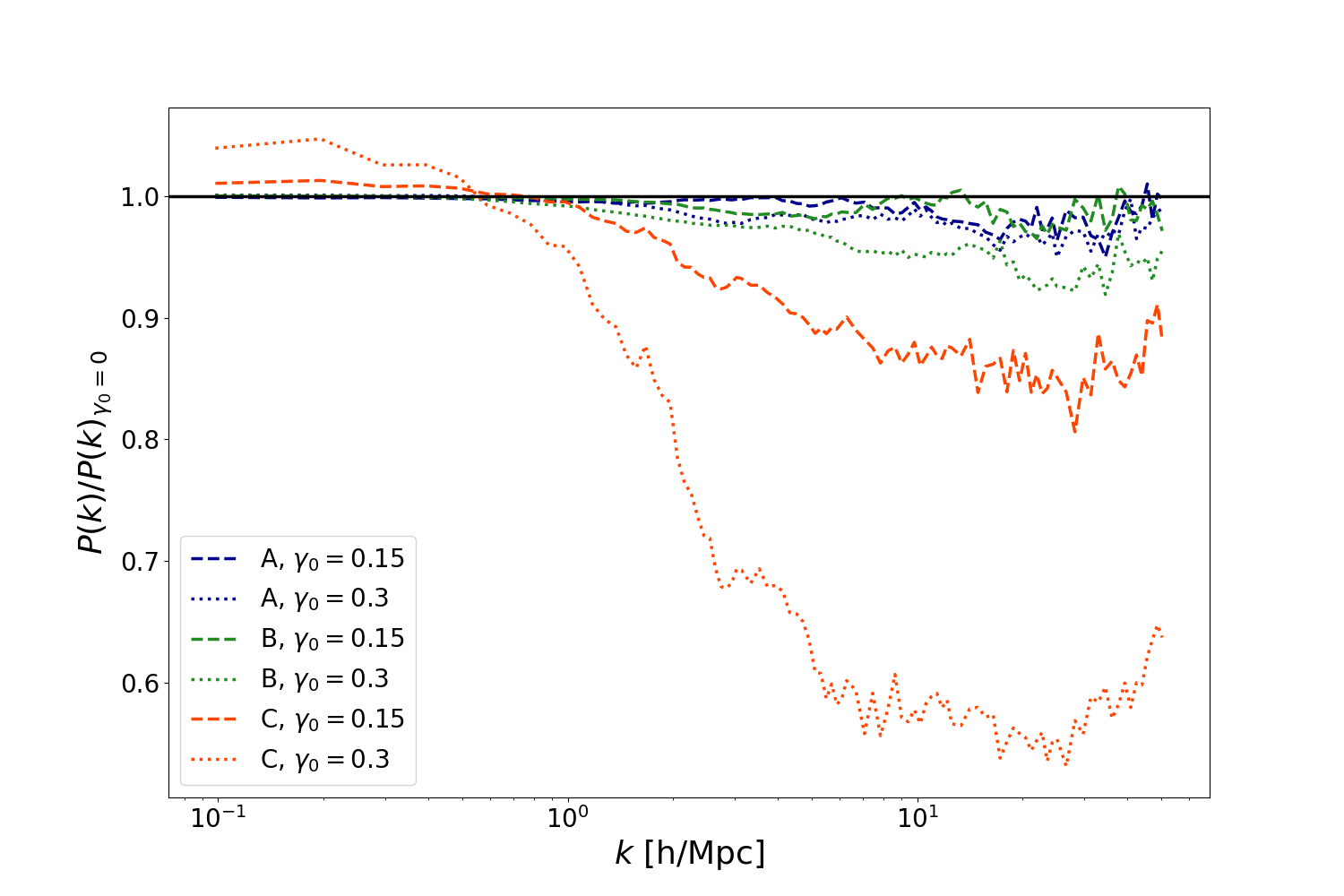}}
  \caption{Ratio of the power spectrum of models A (blue lines), B (green lines) and C (orange lines) with respect to uncoupled case. \label{ps_uncoupled}}
\end{figure}

\begin{figure}
  \centerline{\includegraphics[scale=0.25]{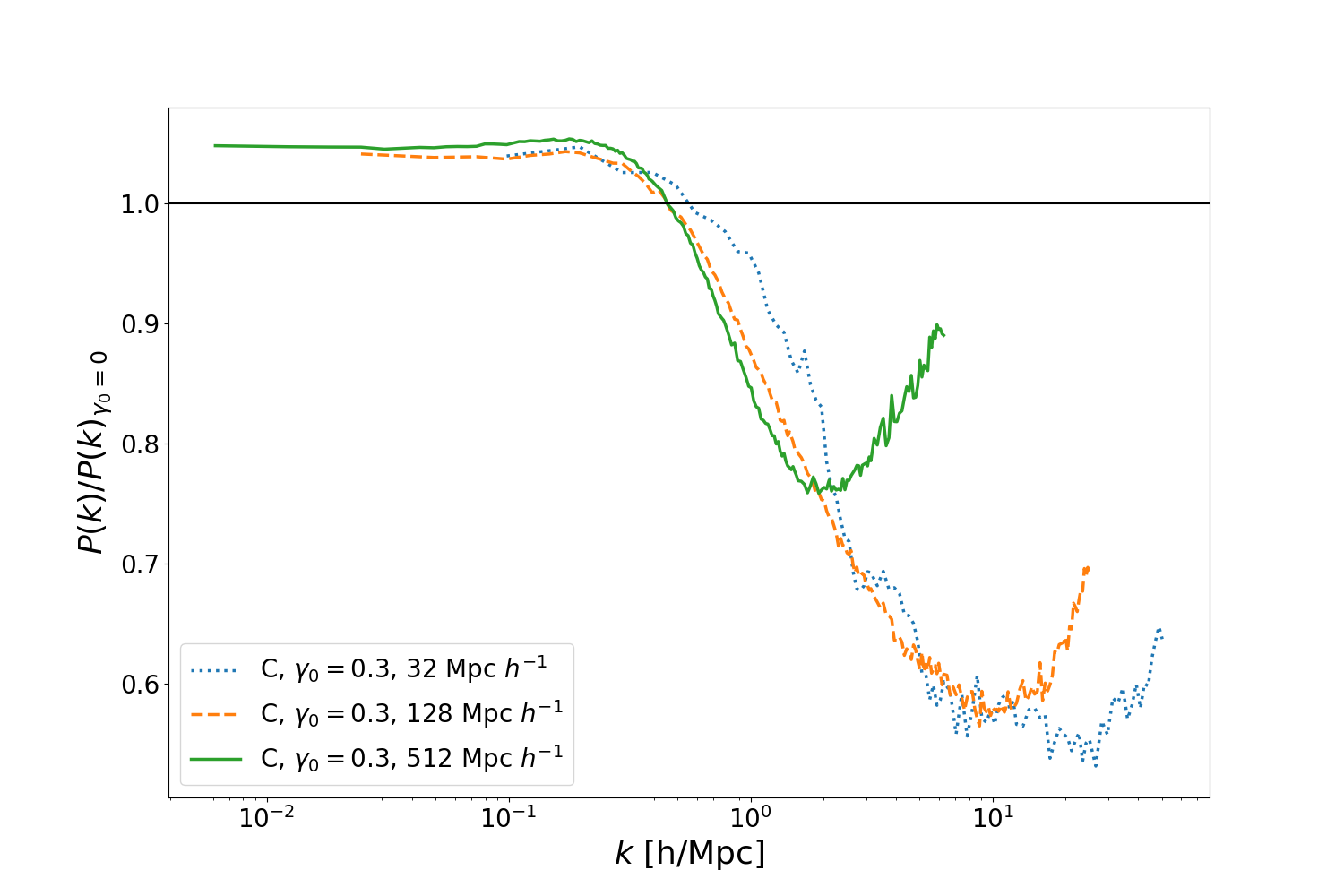}}
  \caption{Ratio of the power spectra of model C (for $\gamma_0 = 0.3$) for all three box sizes considered in this study, with respect to the uncoupled case. \label{ps_uncoupled_bigBox}}
\end{figure}

\begin{figure}

  \centerline{\includegraphics[scale=0.25]{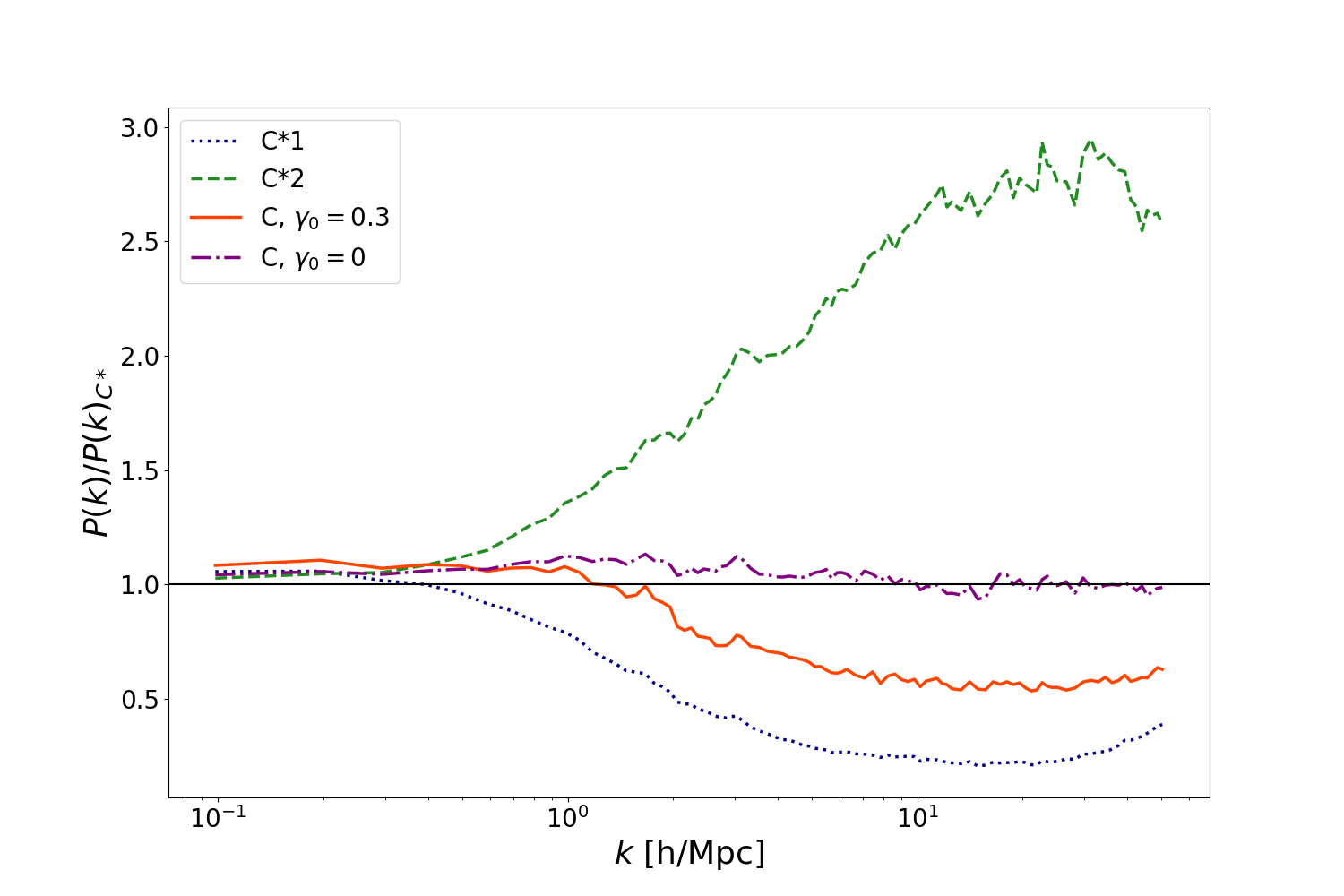}}
  \caption{Ratio of the power spectrum of models C*1 (blue dotted line), C*2 (green dashed line), C with $\gamma_0 = 0$ (orange solid line) and C with $\gamma_0 = 0.3$ (purple dot-dashed line) with respect to model C*. \label{fig:ps_coeffs}}
\end{figure}

To analyse the power spectrum we use POWMES \citep{Colombi_2009}. 
In Figs.~\ref{ps_models} and \ref{ps_uncoupled}, we have plotted the power spectra for all models in the $32$ Mpc h$^{-1}$ box, normalised by the power spectra of $\Lambda$CDM (Fig.~\ref{ps_models}) and the uncoupled models (Fig.~\ref{ps_uncoupled}), i.e. those with $\gamma_0 = 0$, to analyse the effects of the coupling over a wide range of scales. In Fig.~\ref{ps_uncoupled_bigBox} we plot the power spectra for model C with $\gamma_0 = 0.3$ normalised by the same model with no coupling, for all three box sizes considered in this study: $32$ Mpc $h^{-1}$, $128$ Mpc $h^{-1}$ and $512$ Mpc $h^{-1}$. We also plot in Fig.~\ref{fig:ps_coeffs} the power spectra of the models C*1, C*2 and C (all with $\gamma_0 = 0.3$) normalised by the model C* which has $\epsilon_1 = \epsilon_2 = 1$ (i.e. enforcing a standard Euler equation) but the same background evolution as the coupled model.

We can see in Fig.~\ref{ps_models} that, compared to our $\Lambda$CDM model, the quintessence models all have reduced power on all scales, with the difference being of order $15-20\%$ on the largest scales. This arises from the differing background evolution in our quintessence models as compared to that of the $\Lambda$CDM model, as would be expected given the differing Hubble parameters shown in Fig.~\ref{omega_phi_evol}. The fact that this is seen for all values of the coupling tells us that the difference in power spectra is indeed almost entirely due to this modified background evolution: it is not strongly dependent on the coupling. At smaller scales, however, we see a much stronger coupling dependence, especially for model C with large coupling.

If we compare this behaviour with Fig.~\ref{ps_uncoupled}, where we normalise the power spectrum of each model with respect to the uncoupled model of that type, we notice that the power on large scales is enhanced for model C when compared to the uncoupled case, with effectively no change for the other models. Considering the final values of the coefficients of the modified Euler equation given in Table~\ref{tabla_euler_coeff} we would not expect to see large deviations in Fig.~\ref{ps_uncoupled} for models A and B, but would expect to see much more significant deviations for model C. At smaller scales there is again a clear suppression of structure due to the presence of the coupling.

The transition from an enhancement of the structure at larger scales to a reduction at smaller scales seen in Figs.~\ref{ps_uncoupled} and \ref{ps_uncoupled_bigBox} for model C appears to be related to the transition from linear to non-linear scales, as it occurs at a value of $k \sim 0.6$ [h/Mpc]. From Fig.~\ref{ps_uncoupled_bigBox} we can also see that this transition is not dependent on the box size or the resolution. Physically this transition is caused by the reduction of the cosmological friction term as compared to the uncoupled model. For the most extreme case of model C with $\gamma_0 = 0.3$ this is no longer simply a reduction of the cosmological friction but an inversion of the direction of action of the force. Within the linear regime (before shell crossing) the cosmological push is directed parallel to the gravitational accelerations, leading to an enhancement of power at larger scales, whereas in the non-linear regime (after shell crossing) this is no longer the case, causing the cosmological push to generally reduce power on small scales.

These results are consistent with the results of \cite{Baldi&simpson_2015} regarding the dark scattering model, as we will discuss in more detail at the end of this section.


It is important to note, however, that for all our models the coupling parameter $\gamma_0$ appears in the equation of motion for the background quintessence field (\ref{scalar_field_background2}) and so the background evolution is modified when the coupling takes different values. This is also clear from the evolution of $w_{\phi}$ (see Fig.~\ref{eos_models}). Therefore, we fully separate the consequences of the coupling in the Euler equation from the background evolution in Fig.~\ref{fig:ps_coeffs}. In this figure we include model C (orange line), which includes the coupling in both the gravitational force term and the cosmological friction term (with $\gamma_0 = 0.3$), model C*1 (blue dotted line) which includes only the modification to the cosmological friction and model C*2 (green dashed line) which includes only the modification of the gravitational force term. We also include model C without coupling (i.e. with $\gamma_0 = 0$, the purple dot-dashed line). All models are normalised with respect to the model C*, which has the same background evolution as models C ($\gamma_0 = 0.3$), C*1 and C*2, but where the coupling is not included in the Euler equation.

We first note that the enhanced power at larger scales present for models C*1, C*2 and C ($\gamma_0 = 0.3$) is here due entirely to the modified Euler equation, given that the Hubble parameters of all these models are identical. There is enhancement of power at larger scales due to both coefficients $c_1$ and $c_2$, as seen for models C*1 and C*2. The combination of these two effects then leads to a larger total enhancement of power on larger scales, as seen for model C ($\gamma_0 = 0.3$). The comparison of model C ($\gamma_0 = 0$) with model C* (the purple dot-dashed line) allows us to probe the effect of the modified background only. This is because in both of these models the equation of motion of the dark matter is unmodified, but the Hubble parameters differ. This was discussed earlier in the context of Fig.~\ref{omega_phi_evol}. Referencing that figure, we can see that the Hubble parameter for model C* takes \textit{larger} values than those for model C without coupling. We would thus expect a larger contribution of cosmological friction in model C* as compared to model C without coupling, leading to \textit{less} structure at $z=0$ in model C*. Normalising with the power spectrum of this model we thus expect values above unity in Fig.~\ref{fig:ps_coeffs} for the purple dot-dashed line, which is indeed what we see. Thus we have two competing effects in our models: the Hubble parameter is increased (briefly, at late times) when we include the coupling, leading to a reduction in structure as compared to an uncoupled model. The behaviour of the coefficient $c_1$ in the modified Euler equation, however, means that the \textit{effective} Hubble parameter for the dark matter is reduced, which would cause an enhancement of structure. In Fig.~\ref{ps_uncoupled} we have both effects included. The fact that, on larger scales, the coupled models (for model C at least) exhibit enhanced power as compared to the uncoupled models, demonstrates that the effect of the coupling in the modified Euler equation dominates over the modified background evolution.

At smaller scales we see an interesting behaviour: the modified cosmological friction term alone \textit{decreases} structure, while the modified gravitational force term alone \textit{increases} structure. The physical reason for the latter is clear: the effective gravitational force is stronger thus more structure is formed. It should be noted that the scale-dependence of this effect is due to the short timescale over which the gravitational force term becomes modified (see Fig.~\ref{fig:euler_coeffs}). Larger scales have simply not had time to be strongly affected. Turning to the modified cosmological friction term (model C*1), the effect of the very strong coupling in model C ($\gamma_0 = 0.3$) is to change this term from a friction which acts antiparallel to the particle velocities to a force which acts parallel to the velocity vector of the dark matter (due to the sign change in the coefficient $c_1$). Thus the velocities in model C*1 are \textit{enhanced} as compared to the model C*, but the effective gravitational force remains as normal. Dark matter particles are then pushed out of overdensities, resulting in less structure on smaller scales. Interestingly, the combination of these two effects (in model C) still results in a significant reduction of structure at small scales, as the cosmological push overcomes the increased effective gravitational force.


Our results appear to be consistent with the study of \cite{Baldi&simpson_2015} of the dark scattering model, where a momentum transfer between DM and DE leads to a modified cosmological friction term (without a modified gravitational force term). Specifically, we can compare with their case of a constant dark energy equation of state $w = -1.1$. Although the background evolution differs between our models and those of \cite{Baldi&simpson_2015}, the effect on the cosmological friction term is comparable. In their model with $w = -1.1$ the cosmological friction term is reduced as compared to the standard model. They similarly observe an \textit{enhancement} of power as compared to their uncoupled models at large scales with a \textit{reduction} of power at small scales arising purely from a modified cosmological friction term and the modified background evolution. Furthermore, the scale-dependence at small scales and the scale-independence at larger scales is very similar to that seen in our models.

\subsection{Halo properties}
For the analysis of the matter distribution in our simulations, we used the Amiga Halo Finder (AHF) code \citep{Gill_2004, Knollmann_2009}. A velocity criterion is also applied to particles which have been associated to the halo by the density criterion.

Since we are modifying the gravitational force in our models this velocity criterion must also be modified. Thus in AHF we use an effective gravitational constant $\tilde{G}$, which comes from the multiplication of $G$ with the appropriate value of $c_2$ given the redshift under consideration. In Table \ref{tabla_G*} we give this value for $z=0$, the redshift at which all of our results are determined.
\begin{table}
\begin{center}
\begin{tabular}{ l | l l l}
 \hline
 Model & A & B & C \\
 \hline \hline
$\gamma_0$ = 0.15 & 4.319 $\times 10^{-9}$ & 4.333 $\times 10^{-9}$ &  4.765 $\times 10^{-9}$ \\
$\gamma_0$ = 0.3 & 4.366 $\times 10^{-9}$ & 4.419 $\times 10^{-9}$ & 6.594 $\times 10^{-9}$\\
 \hline
\end{tabular}
\caption{The effective gravitational constant $\tilde{G}$ of our models. All these quantities are expressed in Mpc {km}$^2$/$M_{\odot}$s$^2$. \label{tabla_G*}}
\end{center}
\end{table}

Particles associated to a halo due to the density criterion will then be assigned to that halo if their velocity is lower than $1.5$ times the escape velocity, i.e. $v<1.5v_{esc}$, where
\begin{equation}
    \label{eq:binding}
    v_{esc} = \sqrt{2|\Phi|}
\end{equation}
and 
\begin{equation}
    \frac{d\Phi}{dr} = \frac{\tilde{G}M(r)}{r^2}
    \label{G_ahf}
\end{equation}
with $r$ the radius from the halo center, $\Phi$ is the peculiar gravitational potential, $M(r)$ the mass inside the halo and $\tilde{G}$ the rescaled effective gravitational constant. The total number of halos found at z = 0 for each type of model, for differing couplings, is given in Table \ref{Tabla:numero_de_halos}.

\subsubsection{Mass functions}
\begin{figure}
  \centerline{\includegraphics[scale=0.25]{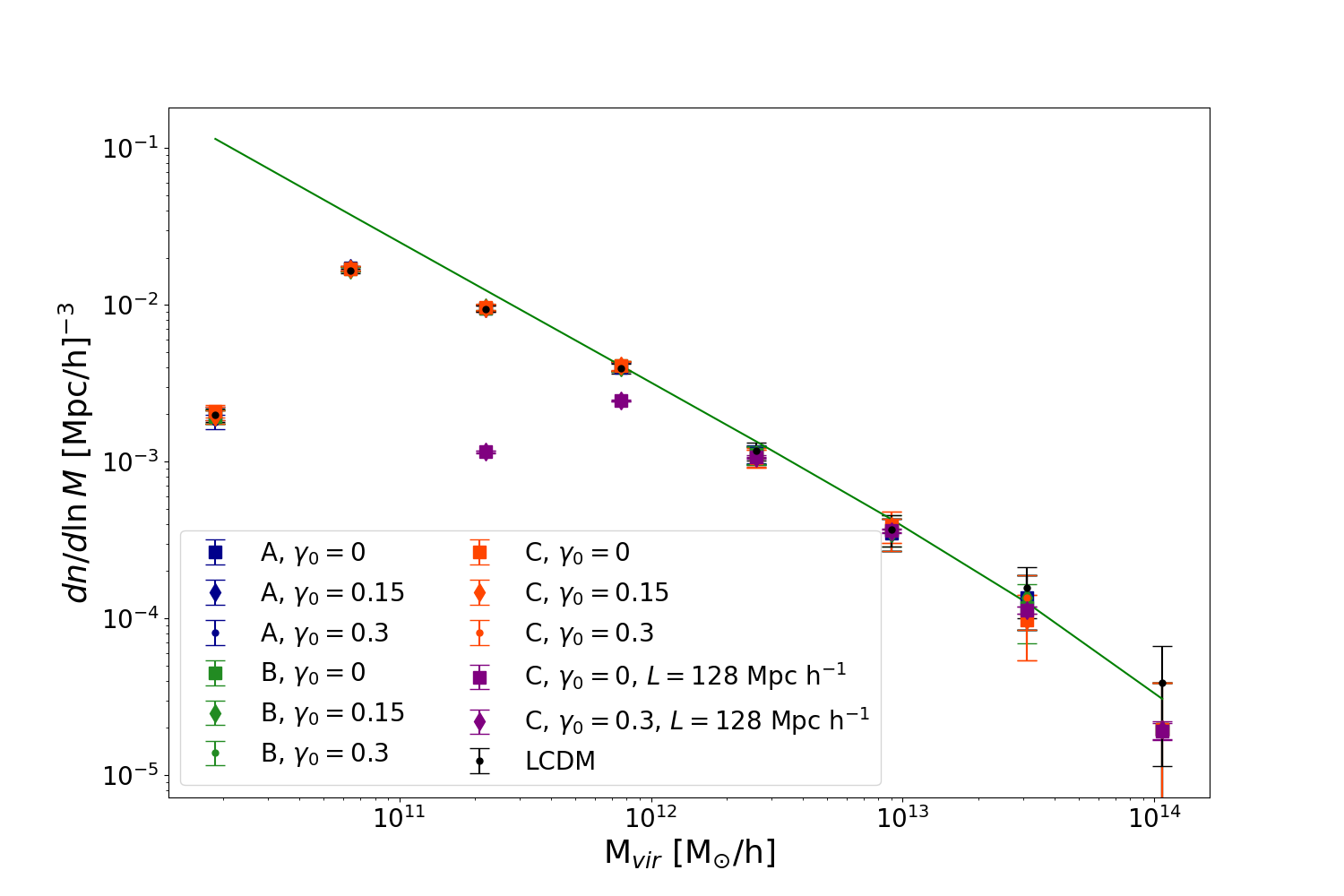}}
  \caption{Halo mass functions for models A (blue symbols), B (green symbols) and C (orange and purple symbols). The green line is the fitting function of \protect\cite{tinker2008}. Poissonian error bars (see text) are also shown.}
  \label{fig:hmf_models}
\end{figure}

We show the halo mass function for all of our models in Fig.~\ref{fig:hmf_models}. We have used the COLOSSUS package \citep{Diemer_2018} to compare our results with an analytical HMF fitting function defined in \cite{tinker2008}. This definition is derived from halos identified in simulations using the spherical overdensity (SO) method, which agrees with the method used by the AHF code.

We have added Poissonian error bars to Fig.~\ref{fig:hmf_models} where the halo number count in each bin is given by $N \pm \sqrt{N}$


Due to the limited volume and mass resolution of our simulations, we can only effectively resolve halos in the mass range of $10^{11}$ - $10^{14}$ $M_{\odot}$ in our smallest box size of $L = 32$ Mpc h$^{-1}$. This leaves us without information about structures with smaller masses ($< 10^{11} M_{\odot}$). At higher mass scales our results are comparable to the reference halo mass function within the error bars. We should caution however that we have only a small sample of massive halos present in the models using the smallest box size. We also show the halo mass function for two models with $L = 128$ Mpc h$^{-1}$. The mass resolution in these models is poor, thus we cannot meaningfully resolve halos below $\sim 10^{12}$ M$_{\odot}$. The error bar in our most massive mass bin is reduced for these models. An increase in the mass resolution of these models would allow us to resolve halos for a wider range of masses. The disagreements with the halo mass function at low and high masses appear to be largely unrelated to the coupling, with reasonable agreement in the HMF for all models in the intermediate mass range.


\begin{table}
\begin{center}
\begin{tabular}{c|ccc}
\hline
$\gamma_0$ & Model A & Model B & Model C \\
\hline \hline 
0 & 1766 & 1755 & 1777\\
0.15 & 1766 & 1746 & 1762\\
0.3 & 1745 & 1755 & 1764\\
\hline
\end{tabular}
\caption{Total number of halos at z = 0 obtained with AHF.}
\label{Tabla:numero_de_halos}
\end{center}
\end{table}

\subsubsection{Density profiles}
To analyse the change in the halo density profile due to the coupling we consider the most massive halo in each of our models, which is of the order of $10^{14}$ M$_{\odot}$. 
Note that, due to the overall similarity in structure across all of our models, the selected halos have similar positions within the computational volume. Specifically, we have verified that the distribution of $x$, $y$ and $z$ coordinates of the centre-of-mass of the most massive halo across all models has a maximum standard deviation of $\sim 75$ kpc h$^{-1}$, which is well within the virial radii of the halos (the minimum $R_{\text{vir}}$ for the most massive halo across all models is $\sim 690$ kpc h$^{-1}$).

\begin{figure*}
  \centerline{\includegraphics[scale=0.35]{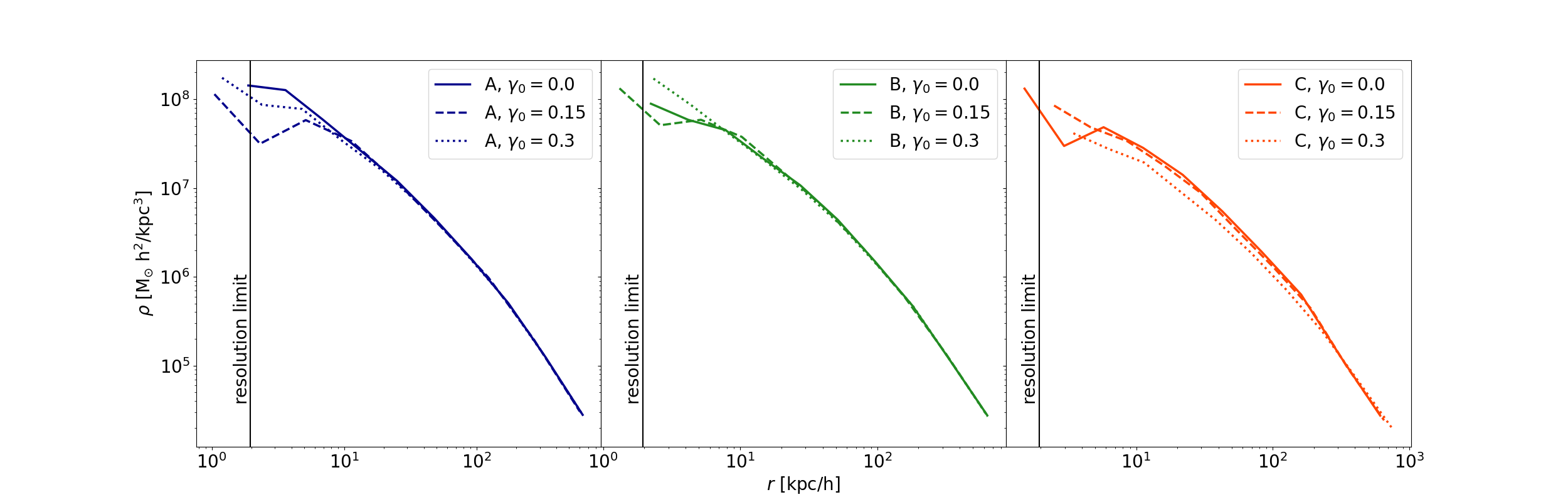}}
\caption{Density profiles for the most massive halo in each model (left panel: model A; middle panel: model B; right panel: model C).}{\label{fig:dens_prof_halo1}}
\end{figure*}

The profiles are shown in Fig.~\ref{fig:dens_prof_halo1} 
separated into three panels: in the left panel we have model A with $\gamma_0 = 0$, 0.15 and 0.3; and the same for model B (middle), and model C (right panel). The black vertical line indicates the resolution limit of our small box simulations, which we have chosen to be given by $4$ grid cells at the highest level of refinement, i.e. $4\Delta x$, where $\Delta x = 1.95$ kpc h$^{-1}$.

Models A and B do not show any difference in their density profiles with a change in the coupling constant. In fact, at radii greater than $\sim 50$ kpc h$^{-1}$, the density profiles for each value of $\gamma_0$ seem to be almost identical. For model C we see that an increase in the coupling constant causes a \textit{decrease} in the inner density of the halos, with this effect extending beyond the very inner regions, such that the profiles are similar only in the outermost parts of the halo. Although there is some variation in mass in the most massive halo in the simulations due to the change in coupling (see Table \ref{Tabla:halos_density_prof}) this is not sufficient to explain the changed density profile.


\begin{table}
\begin{center}
\renewcommand{\arraystretch}{1.2}
\begin{tabular}{ c|cccc }
\hline
Model & $\gamma_0$ & $N_{subs}$ & $N_{part}$ & $M_{halo}$ [$M_{\odot}/h]$\\
\hline \hline
& 0 & 21 & 101880 & $1.319 \times 10^{14}$  \\
A & 0.15 & 21 & 101583 & $1.315 \times 10^{14}$  \\
& 0.3 & 21 & 102019 & $1.321 \times 10^{14}$ \\
\hline
& 0 & 21 & 101163 & $1.310 \times 10^{14}$  \\
B & 0.15 & 21 & 101594 & $1.315 \times 10^{14}$  \\
& 0.3 & 20 & 100927 & $1.306 \times 10^{14}$ \\
\hline
& 0 & 19 & 93438 & $1.210 \times 10^{14}$  \\
C & 0.15 & 19 & 96313 & $1.247 \times 10^{14}$  \\
& 0.3 & 23 & 107877 & $1.396 \times 10^{14}$ \\
\hline
\end{tabular}
\caption{Most massive halos selected for density profile analysis.\label{Tabla:halos_density_prof}}
\end{center}
\end{table}

\begin{figure*}
  \centerline{\includegraphics[scale=0.28]{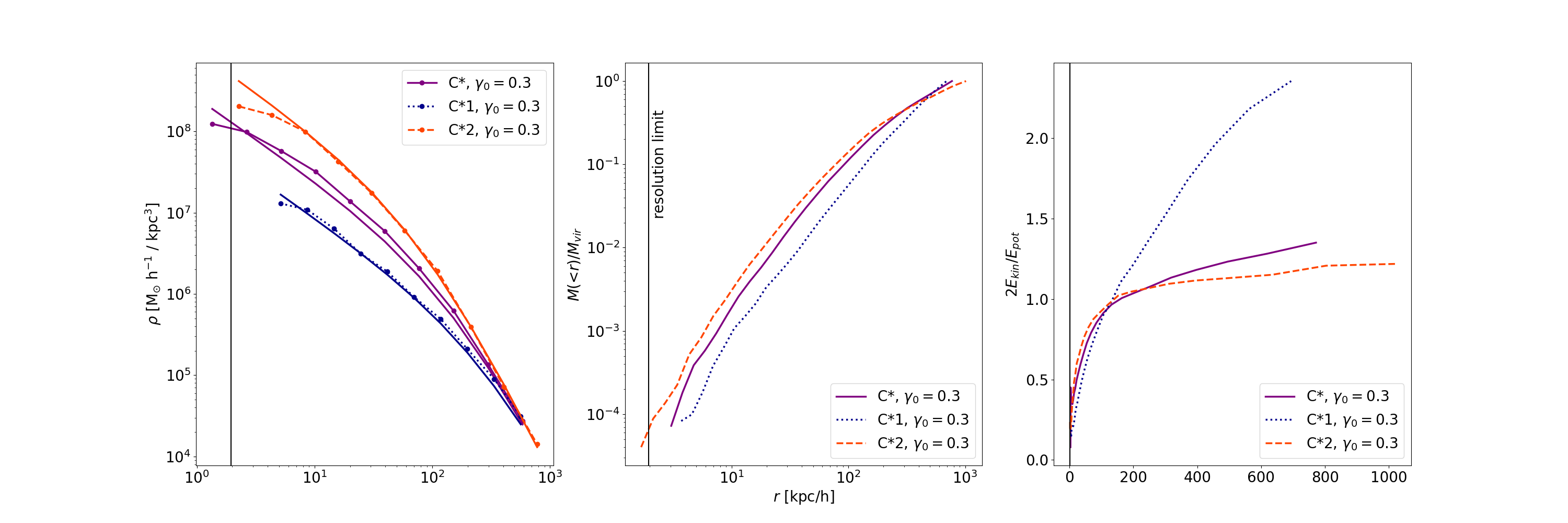}}
  \caption{Profiles for the most massive halos in models C* (no coupling), C*1 (only modified cosmological friction) and C*2 (only modified gravity term). \textit{Left panel:} density profiles. The dotted lines are the simulation halo profiles, while the solid lines are analytic NFW profiles using the virial mass and concentration parameter determined by AHF for each halo. \textit{Middle panel:} cumulative mass profiles. \textit{Right panel:} virial ratio profiles.}
\label{fig:profiles_halo_coefficients}
\end{figure*}

In Fig.~\ref{fig:profiles_halo_coefficients} we show the individual behaviour of each coefficient on the density profiles in the left panel, we show the cumulative mass distribution in the middle panel and the virial ratio in the right panel. We define the virial ratio to be $2E_{\text{kin}}/|E_{\text{pot}}|$, where $E_{\text{kin}}$ is the average kinetic energy within the given radius and $E_{\text{pot}}$ is the average potential energy (as determined by AHF). For this plot we have again selected the most massive halos, this time from the simulations described in Table \ref{tabla_simulaciones_coeffs}. The density profile for the model C*2 (orange solid line) is significantly enhanced in the inner region as compared with the other models. For model C*1 (blue dotted line), the opposite behavior occurs, where a reduction in the innermost regions of the profile is observed. We can also see that there is a lack of particles in the very inner regions of the most massive halo of this model, as compared to the others. In the case of model C* (no coupling) the profile is between these two extremes. We also show in the same figure the fitted NFW halo profiles using the parameters determined by AHF. This is consistent with other works (see e.g. \citealp{Baldi_2010}) that show that a time-dependent enhancement of the gravitational force modifies the virial equilibrium of the halos and leads to an increase in the halo density profile (model C*2). The cosmological push term removes enough dark matter to substantially reduce the halo profile, as compared to model C*. Thus the reduced profiles seen in Fig.~\ref{fig:dens_prof_halo1} are caused by the cosmological push overcoming the effective gravitational attraction within the halo, resulting in a reduction of the inner density. From the middle panel in Fig.~\ref{fig:profiles_halo_coefficients} we can see that the mass distribution is more centrally concentrated in the halo of model C*2, and again we see the reduction of particles in the halo of model C*1, with more of the total mass of the halo being found at larger radii. Finally, in the right panel, the virial ratios for the halos in models C*2 and C* tend towards values reasonably close to unity at large radii indicating that they are close to virialised systems. The halo of model C*1, however, is significantly ``over-virialised'' at large radii, indicating the presence of high kinetic energy particles in the outer region of the halo, deposited there by the effect of the cosmological push. It should be noted that these particles have been identified as being bound to the halo, according to the binding criterion used in Eq.~\ref{eq:binding}, taking into account the enhanced effective gravitational constant of this model.


In summary, for the model which exhibits a significant effect on the halo density profile (model C) we see a \textit{decrease} in the central density. Unfortunately, given the mass resolution of our simulations, we are unable to investigate the density profiles of lower mass halos that would correspond to dwarf galaxies, where the cusp-core problem presents itself. While we hope to improve this in future work, the underlying physical reason for the decreased slope of the density profile seen here should apply equally to all halos, regardless of mass. Thus we would expect less cuspy profiles in these models also for low mass halos.

\subsubsection{Velocity dispersions}
\label{sec:velocity_dispersions}

\begin{figure*}
  \centerline{\includegraphics[scale=0.28]{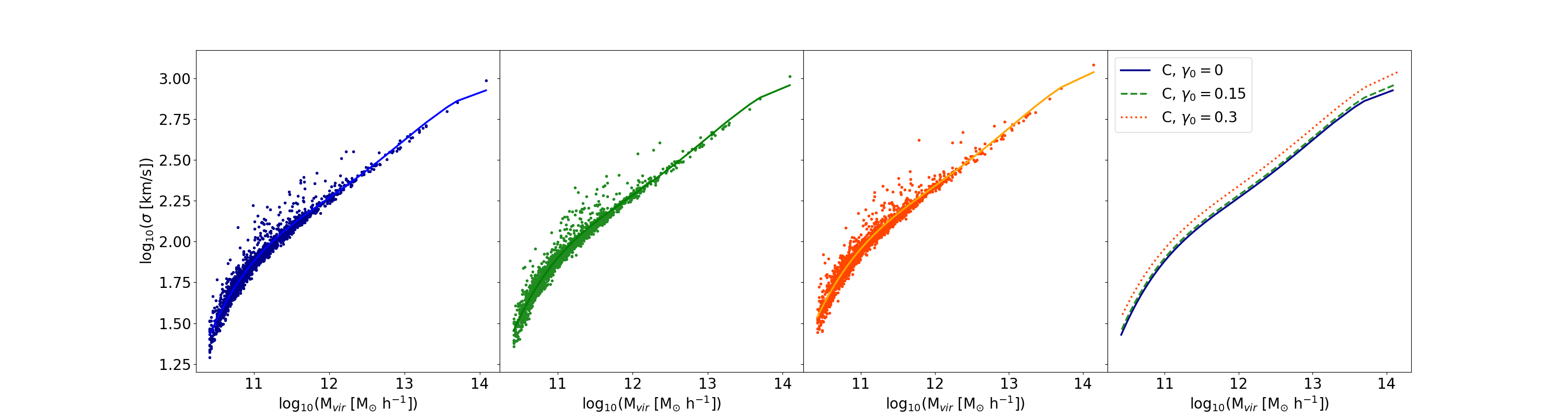}}
\caption{Velocity dispersions of all halos identified in model C for $\gamma_0 = 0$ (\textit{left-most panel}), $\gamma_0 = 0.15$ (\textit{second panel}) and $\gamma_0 = 0.3$ (\textit{third panel}). In all cases a fourth-order polynomial fit line has been added. Each fit line is then directly compared in the plot in the right-most panel.}{\label{fig:sigV}}
\end{figure*}

Given our previous results, showing minimal effects of the coupling for models A and B, we will concentrate mostly on model C for the remainder of our study, except where otherwise indicated. Fig.~\ref{fig:sigV} shows the (logarithm of the) 3D velocity dispersions $\sigma_v$ for all the halos of model C (as determined by AHF) for all values of the coupling. 
The solid lines in each panel of the figure represents a 4th-degree polynomial best-fit line. We compare these best fit lines for different values of $\gamma_0$ in the right panel of Fig.~\ref{fig:sigV}.




We find that the largest velocity dispersions occur for $\gamma_0 = 0.3$ (orange dotted line). The vertical shift compared with the uncoupled case is effectively independent of mass. Given that this is a logarithmic scale, this implies a roughly constant percentage increase in the velocity dispersions as a function of mass. Taking the difference in the best fit lines, we find that the velocity dispersion for halos of mass $10^{11}$, $10^{12}$, $10^{13}$ and $10^{14}$ is increased by $20\%$, $17\%$, $28\%$ and $20\%$ respectively, when comparing the uncoupled model C with the same model and $\gamma_0 = 0.3$. 


Although the cosmological push (the inverted cosmological friction seen in model C with the strongest coupling) appears to reduce the amount of structure at smaller scales (see Fig.~\ref{ps_uncoupled}) and reduce the inner density profile of the halos (see Fig.~\ref{fig:dens_prof_halo1}) the remaining bound material within the halo has a higher velocity dispersion due to the modified virial equilibrium of the halos resulting from the time-dependent enhancement of the effective gravitational force.

These results are again consistent with previous studies of coupled dark matter-dark energy models \cite{Baldi&simpson_2015}, where a similar enhancement of halo velocity dispersion profiles was seen.

\subsubsection{Particle velocity distributions}\label{sec:particle_velocity_distribution}

We have thus far considered the velocity dispersions of the population of dark matter halos in our models, as determined by AHF, and seen the consequences of the coupling in terms of an increased velocity dispersion across all masses, at least for strong coupling in model C. This can be studied in more detail by analysing the full velocity distribution of the particles within the halos. For this, we have selected three halos from model C (considering $\gamma_0 = 0$ and $0.3$) with masses of $\sim 10^{12}$, $10^{13}$ and $10^{14}$ M$_{\odot}$ to see the consequences of the coupling on their constituent particle velocity distributions.

Our selection criteria was the following: we first choose model C with $\gamma_0 = 0$ as our ``base'' model, and identified all halos within this simulation whose masses lay within the intervals $[1 \times 10^n,1.5 \times 10^n]$ where $n$ is chosen to be $n=12$, $13$ or $14$. The most massive halo found within this range is chosen as the ``base'' halo for the comparison (note that the most massive halo in the models under consideration here has a mass of $\sim1.4 \times 10^{14}$ M$_{\odot}$). We then find the halo in the comparison model (model C with $\gamma_0 = 0.3$) whose centre-of-mass coordinates are closest to the ``base'' halo. We finally verify that the chosen halo has a mass that lies within the given mass interval.


\begin{table}
\begin{center}
\renewcommand{\arraystretch}{1.8}
\begin{tabular}{l|ccc}
\hline
Model
& {$10^{12}M_{\odot}$} 
& {$10^{13}M_{\odot}$}
& {$10^{14}M_{\odot}$} \\
\hline \hline
C, $\gamma_0$ = 0 & 133.4 & 251.3 & 564.8  \\
C, $\gamma_0$ = 0.3 & 152.1 & 299.0 & 702.7  \\
\hline
\end{tabular}
\caption{Scale parameter of the Maxwell-Boltzmann distribution functions fitted to the particle velocity distributions of three halos of given mass from model C (see Fig.~\ref{fig:hist_disp_vel_3halos}) \label{tab:scaling_param_modelC_mass_ranges}}
\end{center}
\end{table}



\begin{figure*}
  \centerline{\includegraphics[scale=0.28]{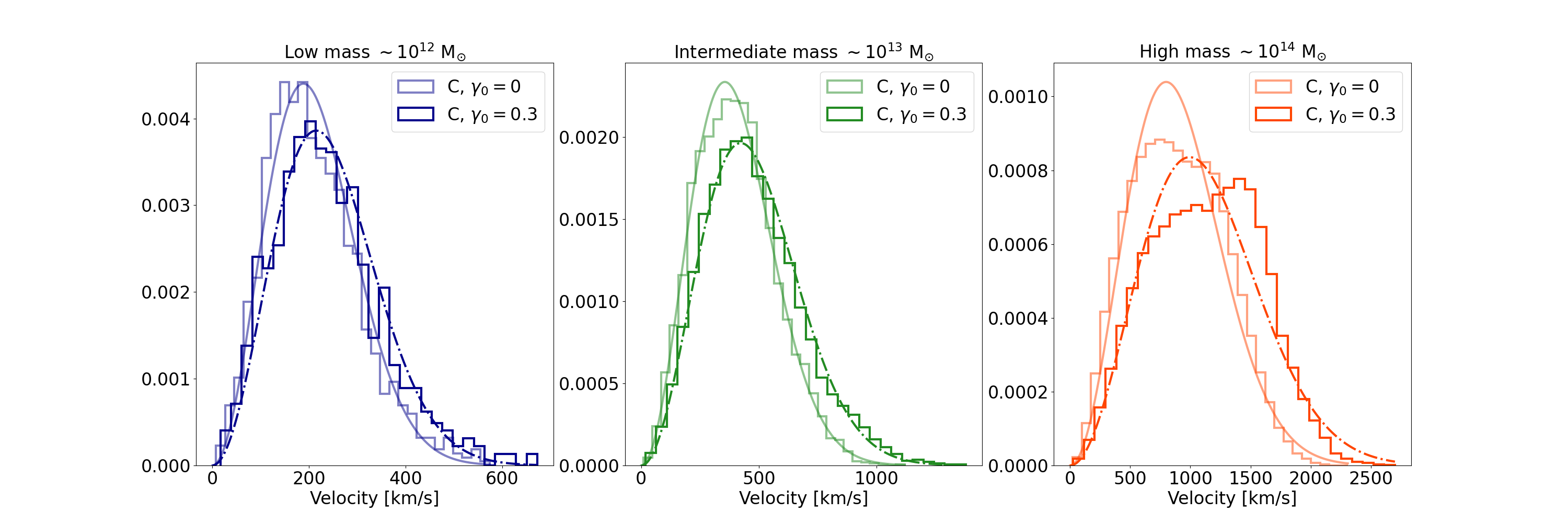}}
\caption{Particle velocity distributions for the low ($\sim 10^{12}$ M$_{\odot}$), intermediate ($\sim 10^{13}$ M$_{\odot}$) and high mass ($\sim 10^{14}$ M$_{\odot}$) halos selected from model C, with $\gamma_0 = 0$ and $\gamma_0 = 0.3$. The curves represent the Maxwell-Boltzmann fit for the uncoupled (solid line) and coupled (dot-dashed line) cases.}{ \label{fig:hist_disp_vel_3halos}}
\end{figure*}

As we can see in Fig.~\ref{fig:hist_disp_vel_3halos}, the velocity distributions for the low mass (left panel), intermediate mass (middle panel) and high mass (right panel) halos are all shifted to higher velocities in the presence of the coupling, with the effect more notable in the high mass halo. We have already seen that the velocity dispersions are increased by the coupling, here we see that the mean velocities are also shifted. The high mass halo appears to have a mean velocity significantly shifted, leading to a skewed distribution.

To quantify the differences between the velocity distributions for different models and different values of the coupling constant, we have fitted Maxwell-Boltzmann (MB) distribution functions to our histograms. While the MB distribution is known to be a rather poor fit to the velocity distributions of gravitational systems, it nevertheless allows us to provide an approximate measure of the change in the distributions due to the coupling. The MB distribution function is given by
\begin{equation}
    f(x) = \sqrt{\frac{2}{\pi}} \frac{x^2}{a^3}  \exp \left( -\frac{x^2}{2a^2} \right)
\end{equation}
where the scale parameter $a$, for an ideal gas, would be given by $a = \sqrt{kT/m}$, where $m$ is the particle mass, $T$ is the temperature and $k$ is the Boltzmann constant. Note that this is a one-parameter distribution function. We can interpret a larger value of $a$ as a ``hotter'' distribution, at least when applying this analysis to the simulation particles that comprise a dark matter halo as the masses are identical. In later sections we will apply a similar approach to the halos themselves. In that case the masses differ, but we will assume that the halo populations in each model are sufficiently similar that the differences in the scale parameter are primarily driven by the effective temperature of the halo distribution.

We have therefore fitted a Maxwell-Boltzmann distribution to each velocity distribution in Fig.~\ref{fig:hist_disp_vel_3halos} and we compare the best-fit values of the scale parameter $a$ in Table \ref{tab:scaling_param_modelC_mass_ranges}. The increase in this parameter for the coupled case in all three mass ranges is evident, indicating that the coupling leads to ``hotter'' velocity distributions.

In selecting these halos we have considered those with similar masses and similar locations within the halo distribution, in order to be able to make a better comparison between each simulation. For the case of the velocity distributions in the left panel of Fig.~\ref{fig:hist_disp_vel_3halos}, the coupled model halo has a mass $10.7\%$ lower than that in the uncoupled model. For the intermediate mass case, the coupled model halo has a mass $9.4\%$ lower, while the high mass halo in the coupled model has a mass $15.7\%$ higher than the uncoupled case.

Although these differences in mass will also lead to differences in the velocity distribution, the effects we see here are considerably larger than expected from the mass difference alone. For the low mass and intermediate mass halos the coupled model halo masses are smaller than those of the uncoupled models, thus we would \textit{lower} mean velocities and dispersions for the coupled model halos if these differences arose purely because of mass differences. In the case of the high mass halo, from dimensional analysis one would expect the moments of the velocity distribution to scale as $\sqrt{M}$, so a $\sim 33\%$ mass difference would translate to a $\sim 12\%$ difference in the moments. We see shifts in the mean velocity in Fig.~\ref{fig:hist_disp_vel_3halos} that are substantially larger than this.

\begin{figure}
  \centerline{\includegraphics[scale=0.25]{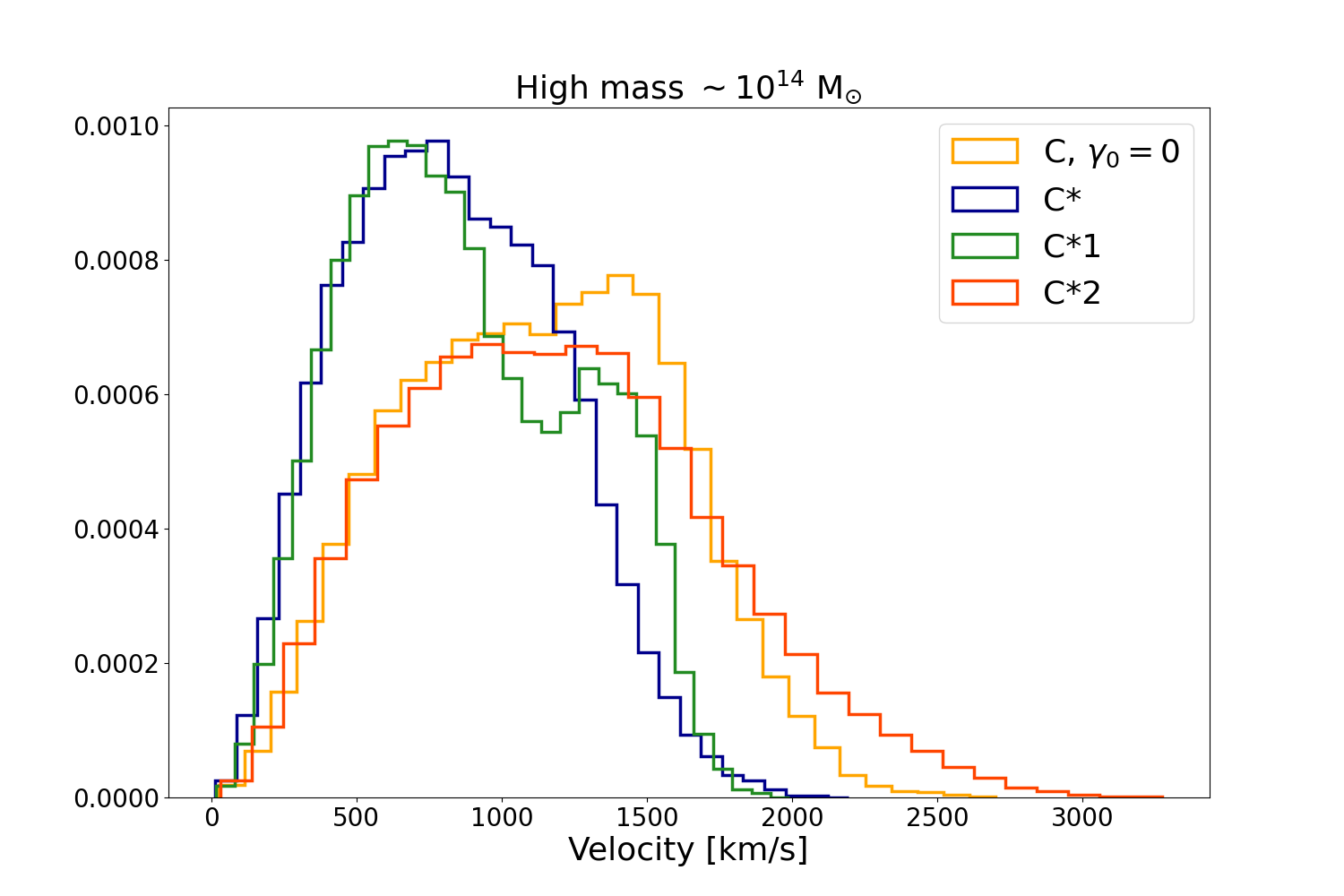}}
\caption{Velocity distribution for the most massive halo selected from the models C with $\gamma_0 = 0.3$, C*, C*1 and C*2.}{ \label{fig:dist_vel_coeffs}}
\end{figure}

As a final analysis of the effect of the modifications to the Euler equation, we have determined the particle velocity distributions for the most massive halo ($\sim 10^{14}$ M$_{\odot}$) from each of the simulations that explore the consequences of each modification separately (and maintain an equivalent background evolution in the uncoupled case). These are the C*, C*1 and C*2 simulations referred to in Table~\ref{tabla_simulaciones_coeffs}. The distributions are shown in Fig.~\ref{fig:dist_vel_coeffs}, compared with the distribution from model C at strong coupling that was already shown in the right panel of Fig.~\ref{fig:hist_disp_vel_3halos}.

\begin{figure*}
  \centerline{\includegraphics[scale=0.25]{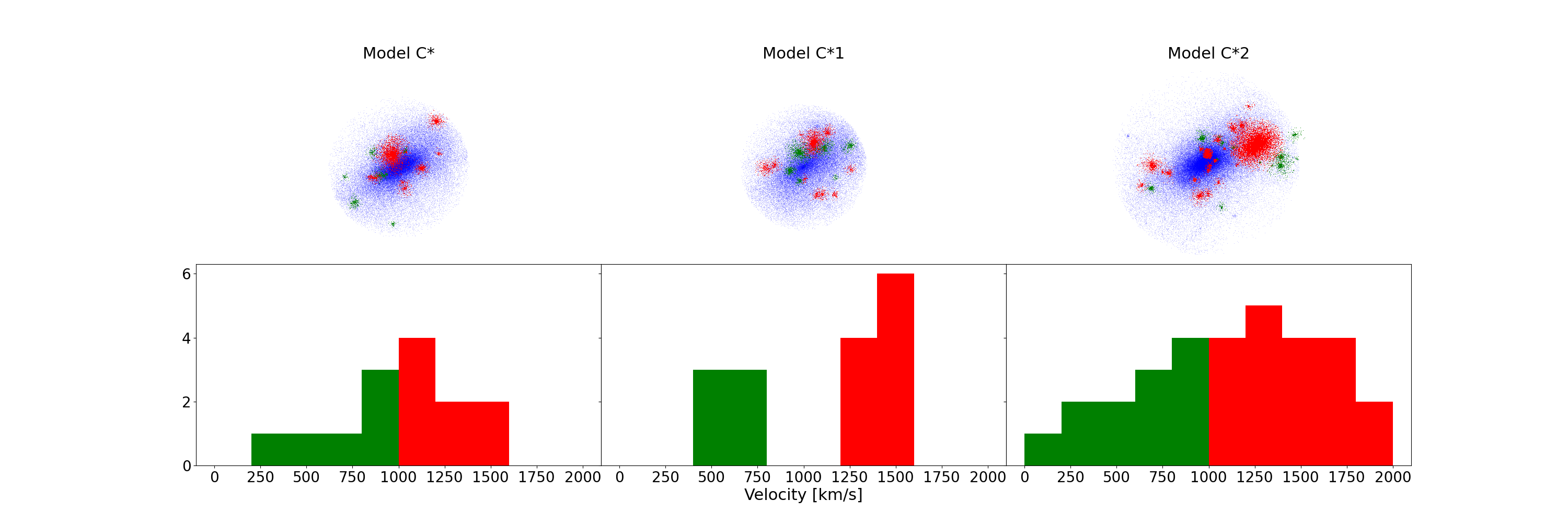}}
\caption{Subhalos within the most massive host halo of models C with $\gamma_0 = 0.3$, C*, C*1 and C*2, with histograms of their velocity magnitudes. Velocities above $1000$ km/s are indicated in red, those below are indicated in green, with the associated halo particles coloured accordingly.}{ \label{fig:subhalos_within_massive_halo}}
\end{figure*}

We find that the velocity distributions of each model show quite considerable differences. Firstly, the distribution for model C* (equivalent background evolution but no coupling at the level of the equations of motion) is peaked at a far lower velocity and has a smaller dispersion than the full model C, as expected. In the case of model C*2 (modified gravitational force term only) the dispersion of the distribution is substantially enhanced, as is the mean velocity (although the distribution is not sharply peaked), compared to model C*. The model containing only the effect of the cosmological push (model C*1) shows the most interesting behaviour, being a bimodal distribution, suggesting the presence of a subgroup of particles with larger velocities that is bound within the halo. The increase in the velocity dispersion in this case is not as pronounced as for model C*2. Thus we can conclude that the distribution seen for the highest mass halo in Fig.~\ref{fig:hist_disp_vel_3halos} (right panel) has an increased mean and dispersion primarily due to the enhancement in the effective gravitational force, and is somewhat skewed (with a very weak bimodality) due to the cosmological push.
The masses of the selected halos from models C*1 and C*2 are $44\%$ lower and $15\%$ higher when compared to the halo selected from model C for $\gamma_0$= 0.3. Again these mass differences alone are not sufficient to explain the differences in the velocity distributions shown here, as discussed earlier.

\subsubsection{Bimodality}
We now investigate further the bimodality exhibited in the particle velocity distribution for model C*1 in Fig.~\ref{fig:dist_vel_coeffs}. In Fig.~\ref{fig:subhalos_within_massive_halo} we show the particle content in the most massive halos of models C*, C*1 and C*2 (projected onto the $xy$ plane), with all plots on the same scale. In the bottom row of this figure we show the velocity distributions of the subhalos within each host. The particles belonging to these subhalos are indicated in the particle plots in the upper row. Those subhalos with velocities (relative to the centre-of-mass velocity of the host) greater than $1000$ km/s are indicated in red, while those with lower velocities are indicated in green. Note that this is simply to aid in the identification of the subhalo populations, we have not applied a velocity cut.

\begin{figure*}
  \centerline{\includegraphics[scale=0.3]{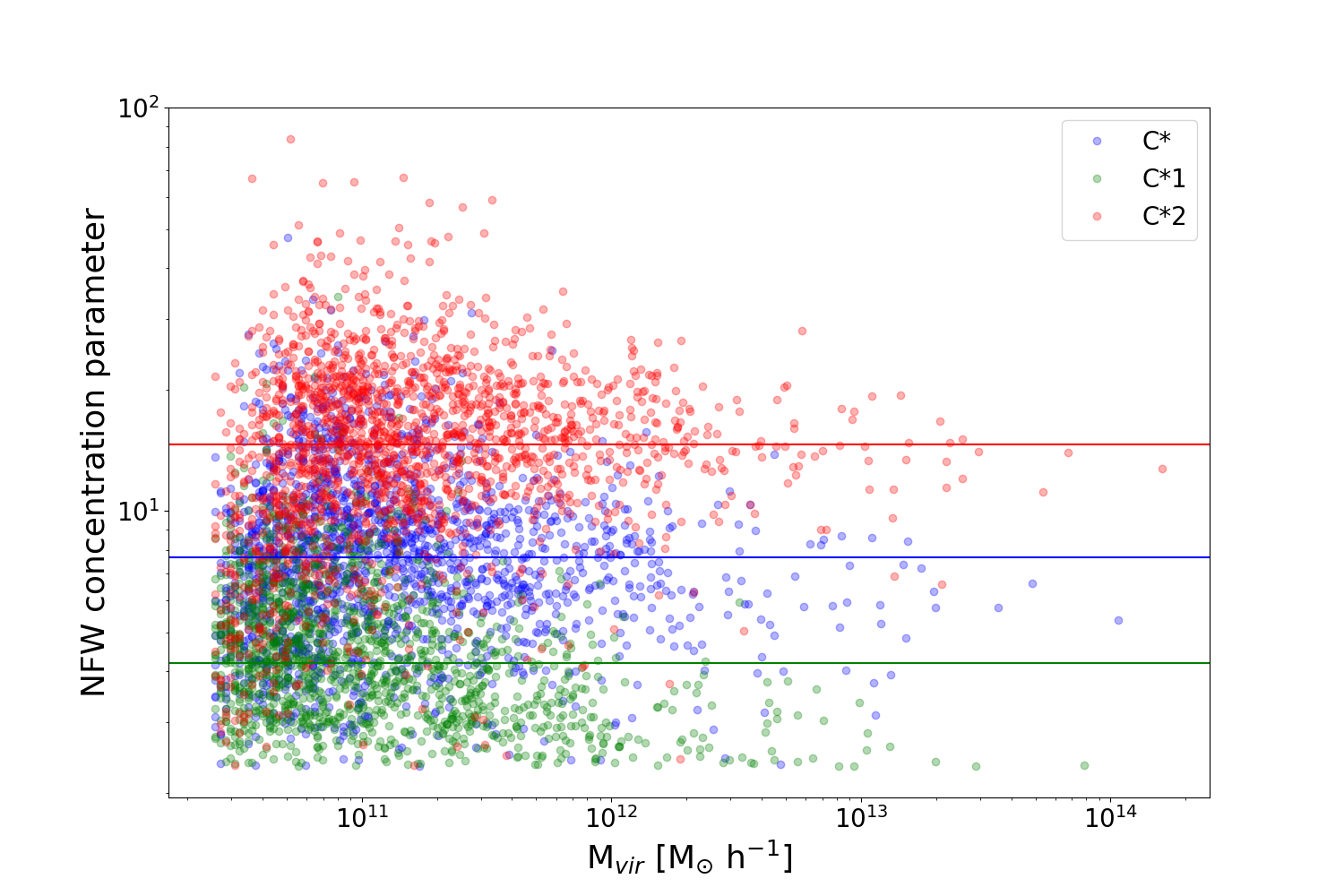}}
\caption{Mass-concentration relation for all halos in models C*, C*1 and C*2. The horizontal lines indicate the median concentration values for each model.}{ \label{fig:mass_conc}}
\end{figure*}

Comparing amongst the models, we can see that the host halo in the models C* and C*1 is less spatially extended than that of model C*2, which is to be expected given the enhanced effective gravitational interaction in the latter model. It is this enhancement which has apparently also led to an increase in the number of substructures within this host, as compared to the other models. Further evidence for this is shown in Fig.~\ref{fig:mass_conc} where we plot the mass-concentration relation for all subhalos within models C* and C*2, clearly demonstrating generally higher concentrations for the same mass in the latter model. Thus the subhalos in model C*2 are likely to be less susceptible to tidal stripping within their hosts.

Turning our attention to the halo velocity distributions, we can see that both models C* and C*2 show continuous distributions across the range of velocities, whereas model C*1 shows a very clear separation between subhalo populations, with one group of fast-moving halos clearly distinct from the other slow-moving group. It is the presence of this group of subhalos that gives rise to the bimodality in Fig.~\ref{fig:dist_vel_coeffs}.

It is clear that structure formation proceeds differently in each model (as evidenced by the differing subhalo populations). In the case of model C*1, we hypothesise that there is an accretion of subhalos whose centre-of-mass velocities diverge from that of their host, presumably due to the influence of the cosmological push term, which is dynamically independent of the accelerations induced from the matter distribution. It may be interesting to look for this dynamical separation of groups within host halos in these kinds of models, perhaps using phase-space analyses. We leave further study of the source of this bimodality for future work.

\subsubsection{Halo velocity distribution}
\label{sec:halo_velocity_distribution}

\begin{figure*}
  \centerline{\includegraphics[scale=0.28]{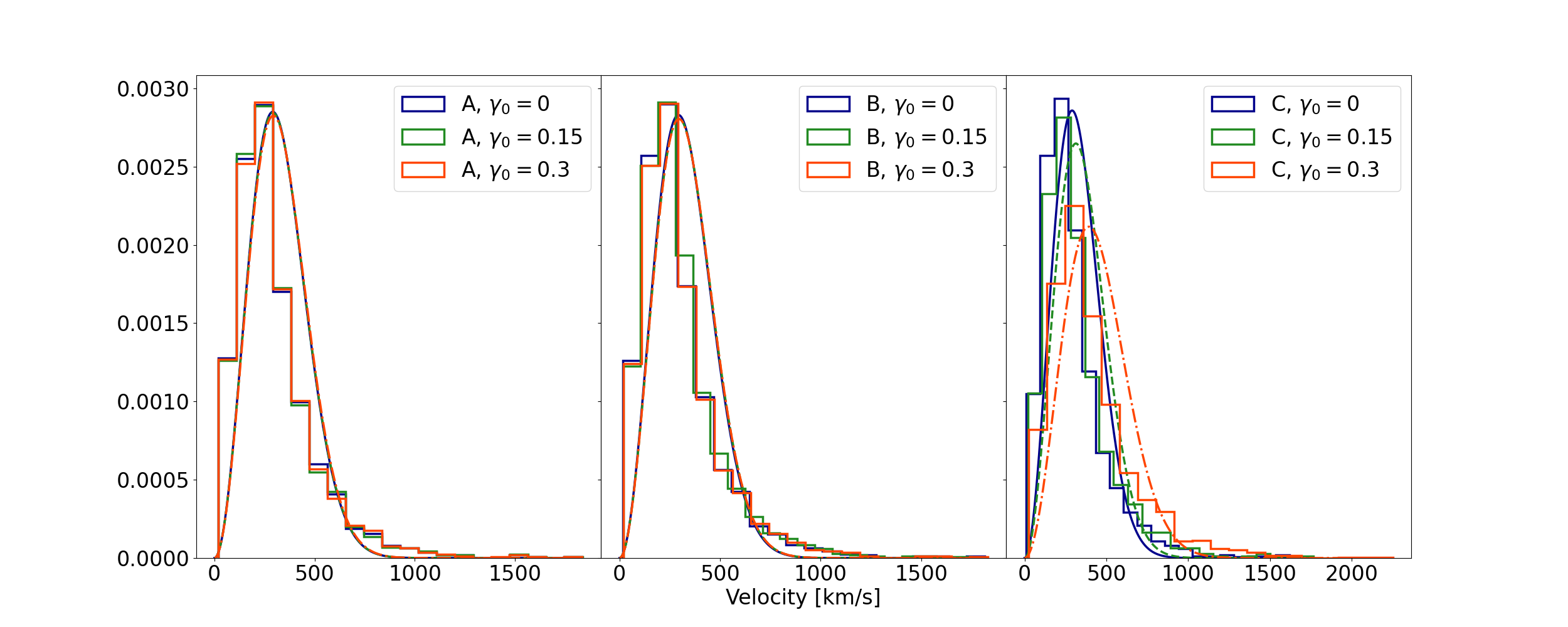}}
\caption{Velocity distributions of all halos for models A, B and C. The curves represent the Maxwell-Boltzmann fit for $\gamma_0 = 0$ (green line), $\gamma_0 = 0.15$ (orange line) and $\gamma_0 = 0.3$ (red line).}{ \label{fig:host_halos_vel_dist}}
\end{figure*}

\begin{figure}
  \centerline{\includegraphics[scale=0.25]{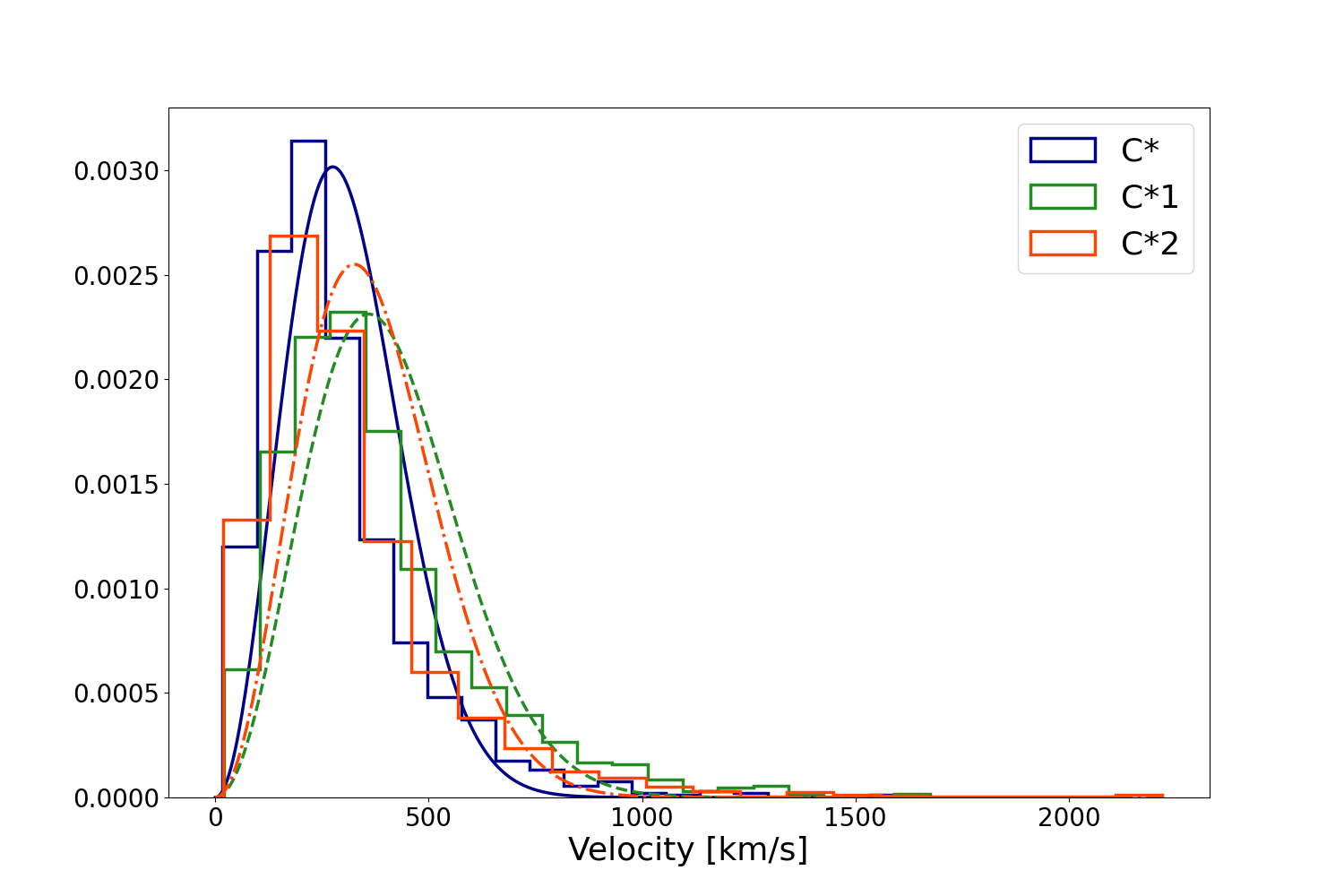}}
\caption{Velocity distributions of all halos in models C*, C*1 and C*2. The curves represent the Maxwell-Boltzmann fit for models C* (green line), C*1 (orange line) and C*2 (blue line).}{ \label{fig:vel_dist_host_euler_case}}
\end{figure}

We have so far analysed the velocity dispersions and velocity distributions of the dark matter within the halos. Now we examine the velocity distributions of the halos themselves. To do so, we have selected all halos from each model, for each value of $\gamma_0$, over the whole mass range. The measured velocity distributions and the fitted Maxwell-Boltzmann distributions for the halos are shown in Figs.~\ref{fig:host_halos_vel_dist} for models A, B and C, and in Fig.~\ref{fig:vel_dist_host_euler_case} for models C*, C*1 and C*2.


\begin{table}
\begin{center}
\begin{tabular}{l|c}
\hline
Model & Scale parameter \\
\hline
A, $\gamma_0 = 0$ & 206.0 \\
A, $\gamma_0 = 0.15$ & 206.5 \\
A, $\gamma_0 = 0.3$ & 207.8 \\
\hline
B, $\gamma_0 = 0$ & 207.4 \\
B, $\gamma_0 = 0.15$ & 209.1 \\
B, $\gamma_0 = 0.3$ & 209.3 \\
\hline
C, $\gamma_0 = 0$ & 205.4 \\
C, $\gamma_0 = 0.15$ & 221.7 \\
C, $\gamma_0 = 0.3$ & 277.4 \\
\hline
C* & 194.7 \\
C*1 & 254.0 \\
C*2 & 230.3 \\
\hline
\end{tabular}
\caption{Scale parameter $a$ obtained from fitting the halo velocity distributions with a Maxwell-Boltzmann distribution for Models A, B, C and the C* models (from Figs.~\ref{fig:host_halos_vel_dist} and \ref{fig:vel_dist_host_euler_case}). \label{tab:scaling_param_halos_subhalos_models_A_B_C}}
\end{center}
\end{table}

As we can see, models A and B show very similar velocity distributions for all halos for all values of the coupling. The same can be seen by looking at the values obtained from the Maxwell-Boltzmann fits (Table \ref{tab:scaling_param_halos_subhalos_models_A_B_C}), where we do not find significant variations for different couplings for models A and B. For our extreme case, model C with $\gamma_0 = 0.3$, we see again a shift in the distribution towards higher mean velocities and larger dispersions. This may be quantified by looking at the Maxwell-Boltzmann fits, where the scale parameter increases significantly for the case of stronger coupling by $35\%$, as compared to the uncoupled model, again showing ``hotter'' velocity distributions.

We now turn to the models that isolate the cosmological push (model C*1) and the modified gravitational force (model C*2). Fig.~\ref{fig:vel_dist_host_euler_case} shows the velocity distributions and their Maxwell-Boltzmann fits for all halos in models C*, C*1, and C*2.


We again see that the cosmological push, model C*1, leads to a shift towards higher velocities and a larger dispersion compared to model C*, as reflected by the values in Table \ref{tab:scaling_param_halos_subhalos_models_A_B_C}. The enhanced gravitational force in model C*2, however, has less effect upon the halo velocities as compared to the cosmological push effect. Although the MB scale parameter for the halos of model C*2 is larger than that of C*1, this appears to be due to a very small number of very high velocity halos shifting the tail of the distribution. The highest velocity halo in model C*1 has $v \sim 1600$ km/s, whereas in model C*2 there are $4$ halos with velocities higher than $2000$ km/s. The main peak of the distribution, however, is broadly comparable with that of model C*, whereas there is clear shift in the peak for model C*1. This suggests that the halo velocities are more affected by the modified cosmological friction, than by the modified gravitational force term, although this latter effect is certainly very relevant for the internal particle distributions within the halos (see Fig.~\ref{fig:dist_vel_coeffs}).



\section{Discussion and conclusions}

In this paper we have analysed structure formation in a coupled DM/DE model, where dark energy arises from a quintessence field, and the coupling is purely at the level of a momentum transfer. Our analysis has been based on numerical N-body simulations using a modified version of the RAMSES cosmological simulations code. We have determined the form of the modified Euler equation in the Newtonian gauge, and then considered the small-scale Newtonian limit. We have shown that the coefficients of the cosmological friction and gravitational force terms in the resulting Euler equation are time-dependent, being functions of the coupling parameter $\gamma_0$, the time derivative of the background quintessence field $\dot{\bar{\phi}}$, the derivative of the potential $V_{\phi}$ and the background dark matter density $\rho$. We have considered exclusively the case where $\gamma_0 > 0$, resulting in a suppression of the cosmological friction term (relative to the standard case) and an enhancement of the effective gravitational force.

After implementation of the modified Euler equation into RAMSES, we have then investigated the consequences for structure formation at recent times, focussing on the power spectra and dark matter halo properties, for three choices of scalar field potential. For two of these potentials (referred to as models A and B) the background evolution is very similar to that of $\Lambda$CDM, with only a small deviation from $-1$ at very late times in the value of the dark energy equation of state parameter $w$. In one of our models, referred to as model C, we have a more pronounced deviation in the background evolution, corresponding to a larger deviation from $w=-1$. This corresponds to a larger contribution from the scalar field kinetic term, leading to non-negligible deviations from unity in the Euler equation coefficients.

Our results demonstrate that, for these models, both the modification of the cosmological friction term and the modification of the gravitational force have an impact in modifying the evolution of structure, especially in model C with the largest coupling. Our specific results are:

\begin{itemize}
\item The power spectrum shows a reduction of structure on small scales for all models, and an increase in structure at large scales for model C.
\item For strong coupling the cosmological push leads to a reduced inner density profile in at least our most massive halos.
\item The mean velocity and velocity dispersion (in model C) of all halos is increased due to the combined effects of the enhanced gravitational interaction and the cosmological push.
\item For this same reason, the particle velocities within the massive halos are also substantially higher with the coupling (in model C) than without, caused by the combination of the effective gravitational force and the cosmological push.
\item We have found a bimodality in the velocity distribution of the most massive halo in our model C*1 that isolates the effect of the cosmological push. We leave for future work further exploration of the exact source of this bimodality. In addition, with higher resolution simulations and improved statistics we hope to study the frequency of this effect in these models.
\item The halo velocity distribution in our models with coupling appears to be primarily affected by the modified cosmological friction term, rather than by the modified gravitational force term.
\end{itemize}


It is worth noting that our study differs from previous work done on simulations of a momentum transfer coupling between dark matter and dark energy \citep{Baldi&simpson_2015, Baldi&simpson_2017}. The models studied here are specific realisations of quintessence scalar field dark energy models and furthermore the modified gravitational force term in our models is non-existent in the dark scattering model. It is also important to note, however, that given the dominant importance of the modified cosmological friction term in our models, our results appear to be very consistent with those reported in \citep{Baldi&simpson_2015, Baldi&simpson_2017}.

In summary, our results suggest that, as far as non-linear structure formation is concerned, most coupled models differ slightly, in general, from their uncoupled counterparts. In the case where the DE equation of state parameter deviates appreciably from $w=-1$ we have a sufficient contribution from the kinetic energy of the quintessence field to generate a substantial additional force upon the dark matter, modifying the evolution of the structure. The most notable modification that results is that of less structure on small scales, and an associated reduction in the internal density profile of the halos, due to the interplay between the enhanced gravitational force acting upon the DM and the so-called cosmological push. This implies that our models, with a positive coupling parameter, could in fact help alleviate some of the tensions present at small scales for $\Lambda$CDM, such as the cusp-core problem. Given the increase in structure at linear scales in our models, it would appear that the $\sigma_8$ tension would not be alleviated in our models, although structure formation at smaller scales can be suppressed. Such a possibility was explored at the linear level in \cite{Pourtsidou_2016} with a negative value for the coupling constant. In our study, the coupling constant is positive for all models.

We should reiterate that we have obtained significant effects in model C because of the similar magnitudes of the two terms in the denominators in equation~(\ref{h_vals}). As discussed in Section \ref{section_modified_euler_eq}, in cases where $\rho_0 \approx 2a\gamma_0 \dot{\bar{\phi}}^2$ we expect substantial deviations, as compared to the standard case, in the coefficients of the Euler equation. If we in fact have an equality in this relationship, we will find singular behaviour in these coefficients. This strongly suggest a limitation in the physical viability of these models, at least for positive coupling.

An interesting aspect of this work that would benefit from more investigation is the combination of a modified dynamics for dark matter with standard dynamics for the baryons. In particular, it would be of interest to explore the possible consequences for dynamical friction, given that the strength of this effect on an object falling into a dark matter halo depends on the velocities of the dark matter particle field, as well as the effective gravitational force. For the dark matter, the effective gravitational force is enhanced by the coupling (increasing dynamical friction) while the particle velocities are also enhanced (decreasing dynamical friction). The additional contribution from the cosmological push to further accelerate the dark matter particles could possibly result in a net reduction in dynamical friction. For baryons, however, the gravitational force is not enhanced, thus one might expect a reduction of dynamical friction for baryons as compared to the uncoupled case. It would certainly be of interest to explore the consequences of this for galaxy dynamics, such as in galaxy mergers and the evolution of bars. This would imply the necessity for hydrodynamical simulations, where the equation of motion of the hydrodynamical material would, of course, be unaffected by the coupling between DM and DE. Alternatively, a population of collisionless star particles could be included, and identified as such within the simulation to allow for their dynamical evolution to be unmodified by the coupling.

A promising avenue for future research in this topic would be to repeat our analysis for $\gamma_0 < 0$. This has already been shown, at the linear level as mentioned earlier, to reduce some tensions in the standard model, specifically with $\sigma_8$ \citep{Pourtsidou_2016}. It is straightforward to consider the evolution of the coefficients $c_1$ and $c_2$ for the case of model C with $\gamma_0 = -0.3$. 
In this case the cosmological friction would now be enhanced but the effective gravitational force is reduced. We have confirmed that the amplitudes of these variations, however, are considerably smaller than seen for the $\gamma_0 = 0.3$ model. This is because, due to the negative $\gamma_0$ in the denominators of equation~(\ref{h_vals}), the singular behaviour discussed earlier cannot arise. For $\gamma_0 > 0$, a strong coupling problem restricts the range to $\gamma_0 < 1/2$. There is no known restriction for negative values of $\gamma_0$, thus a larger absolute value of $\gamma_0$ could potentially be considered in that case. Given that the effective gravitational force is weaker for $\gamma_0 < 0$, this is also likely to reduce the amount of structure formed and perhaps the slopes of the inner densities of that structure. Future studies of these models would also benefit enormously from increased spatial and mass resolution, as well as larger box sizes to explore consequences at very large scales and compare with analytic (linear) perturbation theory results.


\section*{Acknowledgements}

The authors wish to thank the anonymous referee for extremely useful comments which have significantly improved the paper. We also wish to thank Alkistis Pourtsidou for very helpful discussions. The authors acknowledge financial support from FONDECYT Regular No. 1181708. DP thanks Greco Pe\~na for useful discussions and acknowledges the Postgrado en Astrof\'isica program of the Instituto de F\'isica y Astronom\'ia of the Universidad de Valpara\'iso for funding.

\section*{Data Availability}

The data underlying this article will be shared on reasonable request to the corresponding author.



\bibliographystyle{mnras}
\bibliography{mibiblio} 





\bsp	
\label{lastpage}
\end{document}